\documentclass[lettersize,journal]{IEEEtran}
\usepackage{amsmath,amsfonts}
\usepackage{algorithmic}
\usepackage{algorithm}
\usepackage{multirow}
\usepackage{array}
\usepackage[caption=false,font=normalsize,labelfont=sf,textfont=sf]{subfig}
\usepackage{textcomp}
\usepackage{threeparttable}
\usepackage{booktabs}
\usepackage{stfloats}
\usepackage{url}
\usepackage{color}
\usepackage{verbatim}
\usepackage{graphicx}
\usepackage{cite}
\usepackage{diagbox}
\begin{document}

\title{Adaptive Bit Rate Control in Semantic Communication with Incremental Knowledge-based HARQ}

\author{\IEEEauthorblockN{Qingyang~Zhou,~Rongpeng Li, Zhifeng Zhao, Yong Xiao, and Honggang Zhang}

\thanks{This work is accepted by IEEE Open Journal of the Communications Society (OJCOMS).

   Q. Zhou and R. Li are with the College of Information Science and Electronic Engineering, Zhejiang University, Hangzhou 310027, China (e-mail: \{22060638, lirongpeng\}@zju.edu.cn).

   Z. Zhao and H. Zhang are with Zhejiang Lab, Hangzhou, China as well as the College of Information Science and Electronic Engineering, Zhejiang University, Hangzhou 310027, China (e-mail: zhaozf@zhejianglab.com, honggangzhang@zju.edu.cn).
   
   Y. Xiao is with the School of Electronic Information and
   Communications at the Huazhong University of Science and Technology,
   Wuhan, China 430074 (e-mail: yongxiao@hust.edu.cn). Y. Xiao
   is also with Pazhou Lab, Guangzhou, China.

  }

}

\maketitle

\begin{abstract}
Semantic communication has witnessed a great progress with the development of natural language processing (NLP) and deep learning (DL).
Although existing semantic communication technologies can effectively reduce errors in semantic interpretation, most of these solutions adopt a fixed bit length structure, along with a rigid transmission scheme, which is inefficient and lacks scalability when faced with different meanings and signal-to-noise ratio (SNR) conditions. In this paper, we explore the impact of adaptive bit lengths on semantic coding (SC) under various channel conditions. 
First, we propose progressive semantic hybrid automatic repeat request (HARQ) schemes that utilize incremental knowledge (IK) to simultaneously reduce the communication cost and semantic error. 
On top of this, we design a novel semantic encoding solution with multi-bit length selection. 
In this fashion, 
the transmitter employs a policy network to decide the appropriate coding rate, so as to secure the correct information delivery at the cost of minimal bits. 
Moreover, a specific denoiser is further introduced to reduce the semantic errors encountered in the transmission process according to the semantic characteristics of context.
Extensive simulation results have been conducted to verify the effectiveness of the proposed solution. 
\end{abstract}


\section{Introduction}
\IEEEPARstart{W}{ith}  the ever-growing development of artificial intelligence, natural language processing (NLP) and other supporting technologies, 
semantic communication \cite{[28],[29],[30],[34]} raises a surging interest beyond classical reliable communication \cite{[1]}. 
As for classical communication,
according to Shannon’s separation theorem \cite{[31]}, most modern communication systems adopt separate source and channel coding for wireless communication. However,  Shannon's hypothesis suffers in practical situations like finite-length codeword, non-ergodic source or channel distributions.
Therefore, 
semantic communication systems generally adopt advanced joint source channel coding (JSCC) and have manifested the advantages to transmit different types of contents     (e.g. image \cite{[23],[24],[25],[26],[27],[40]}, speech \cite{[22]}, video \cite{[21],[39]}) in a semantic manner.

In the aspect of semantic transmission,
\cite{[2]} proposed to use Bidirectional  Long Short-Term Memory (Bi-LSTM) \cite{[14]} to model  the transmission of sentence in an erasure channel.
The semantic transceiver in \cite{[3]}, which is based on transformer \cite{[13]}, exploited the potential of semantic communication for sentence transmission by combining deep learning (DL)-based JSCC and the NLP technique.
By taking account of  quantization and network compression, \cite{[4]} further explored the possibility of semantic communication for the Internet of Things. 
In \cite{[35]}, semantic communication is extended to multi-user environment.
\cite{[5]} proposed to use reinforcement learning (RL) to optimize the semantic  transmission.
A universal semantic communication system is developed in \cite{[32]} by optimizing the semantic similarity
in an RL manner.
\cite{[33]} introduced a novel  confidence-based  JSCC scheme for semantic transmission.
By leveraging the conventional Reed-Solomon (RS) channel coding and  semantic coding for semantic network, \cite{[6]} exemplified a new direction of semantic communication.
Nevertheless, there still remains some  challenges for semantic transmission system.
In particular, all these methods do not deeply involve the problem of adaptive semantic bit rate control. Furthermore, how to use the minimum communication cost to accurately convey semantic information is worthy of further research. 

In traditional communication, the bit rate represents the proportion of source coding in total coding length. 
However, because semantic communication uses a semantic vector in a JSCC method for information transmission, the adaptive bit rate control in semantic communication more refers to changing the bit length used to represent the semantic vector.
Existing semantic communication  solutions (e.g. \cite{[3],[6]} ) have proved that DL-based coding schemes  yield better performance with more bits used for representing semantic vectors.
When the channel conditions are extremely harsh or the amount of semantic information is large, it is worth using more bits for the success of transmission.
However, because the performance is upper-bounded, when the channel conditions are relatively good or the amount of semantic information is limited, even using comparatively fewer bits, 
a satisfactory result can still be obtained.
But the existing adaptive bit rate control schemes  \cite{[8],[9],[10],[11],[12]} in transmission-level communications ignore the impact of semantic differences in various sentences over changing communication channel conditions.
Hence, there is room for improvement with adaptive bit rate control to overcome the limits of the traditional  fixed bit length solutions.
Building on these overpoints, we preset a new
semantic communication system that can flexibly deal with the semantic differences and varying conditions. 

\begin{figure*}[t]
\centering
\begin{minipage}{0.49\textwidth}
\includegraphics[scale=0.3]{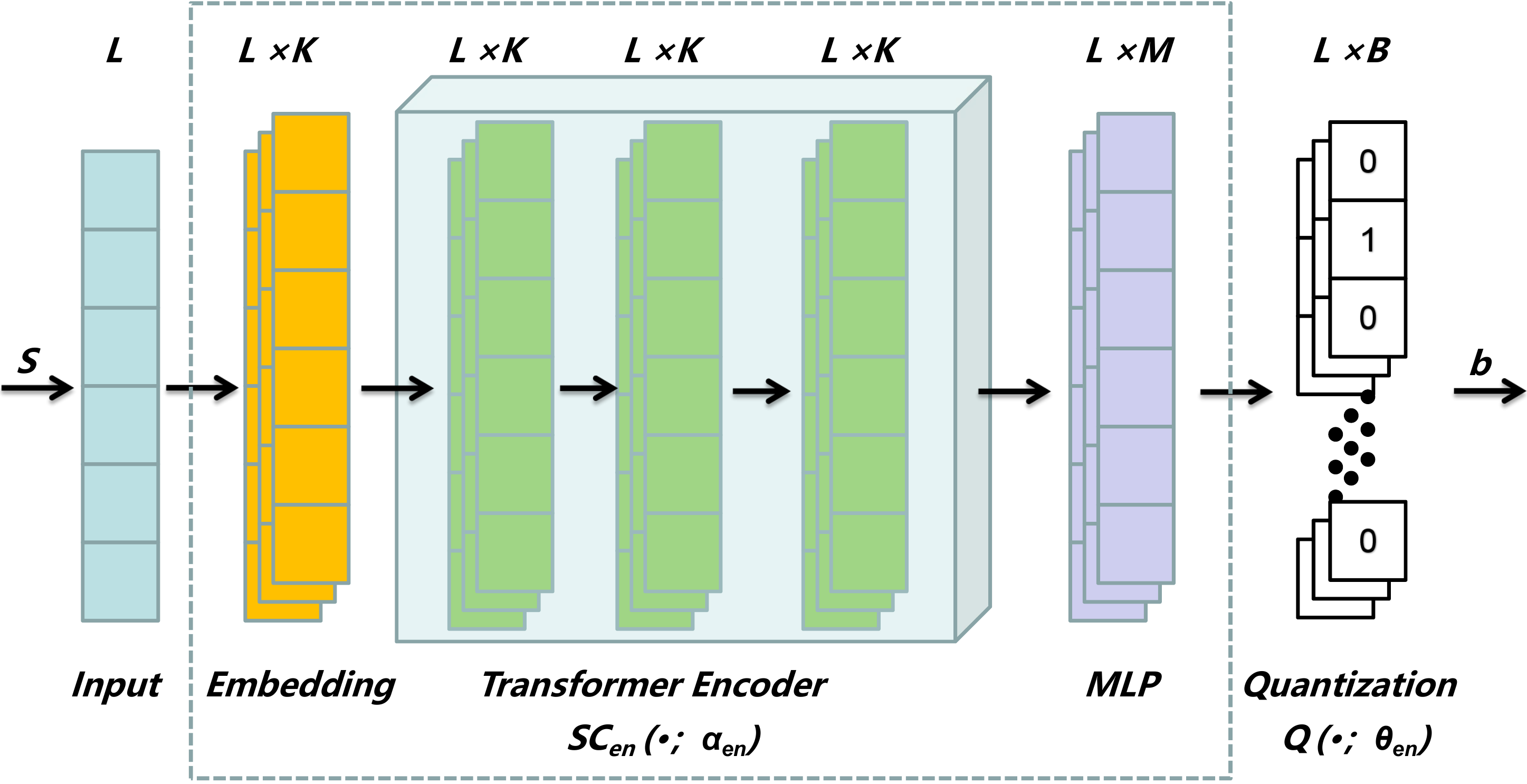}
\end{minipage}
\begin{minipage}{0.49\textwidth}
\includegraphics[scale=0.3]{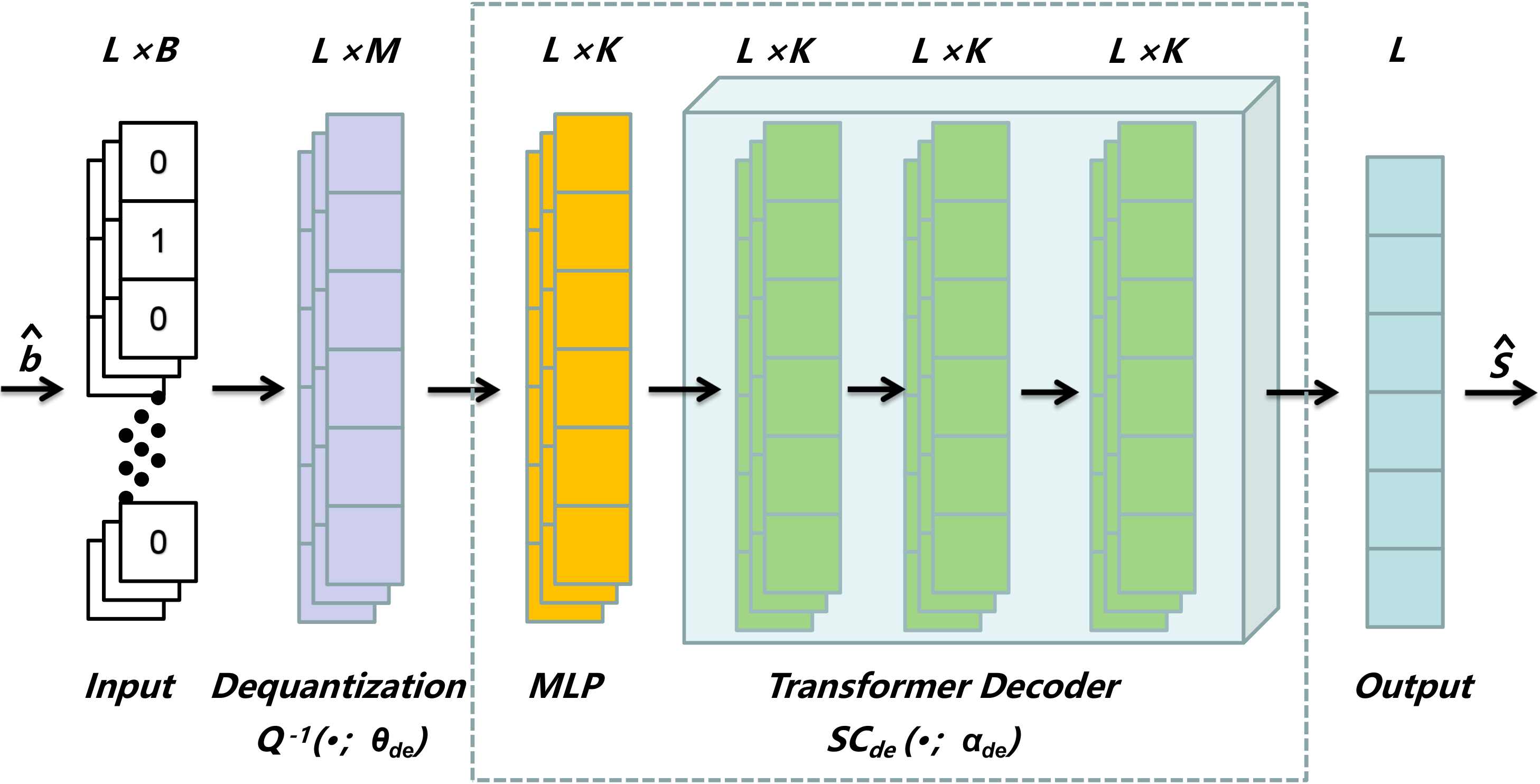}
\end{minipage}
\caption{The basic model for the semantic communication system used in this paper.}
\end{figure*}

In order to achieve the adaptive bit rate control with minimal cost in a lossy environment, a natural solution with enhanced utilization of the retransmissions arises since the sentences failing to pass the error detection standard (e.g. cyclic redundancy check (CRC)) may still contain useful semantic information. 
Adding these incremental knowledge (IK) may contribute to improving the performance of semantic transmission.
In \cite{[6]}, authors  firstly propose to utilize IK through multiple independent coders, which inevitably becomes too cumbersome for practical use.
To overcome it, 
we put forward a novel semantic HARQ (hybrid automatic repeat request) for semantic  transmission, namely incremental knowledge-based HARQ (IK-HARQ).
Based on the existing JSCC scheme \cite{[33]},
we retain the encoder but adjust the decoder to meet the retransmission requirements. 
Two different decoder schemes (i.e., a multi-decoder-based scheme and a single-decoder-based one) are proposed by effectively utilizing IK in cases where HARQ is activated.

Besides considering the number of transmitted bits, the representation form of a semantic vector is also closely related to its 
anti-noise ability.
SNR-adaptive coder \cite{[8],[15],[36]} which makes full use of the SNR information to strengthen the denoising capability has been proposed by various scholars.
However, its practical use in a rapidly changing environment is rather limited, as it needs a relatively accurate measurement of the SNR as a prerequisite. Therefore, we design a novel self-attention based denoising scheme called self-adaptive denoising, which uses the semantic information of the transmission vector rather than the SNR information.

In this paper, we investigate an end-to-end design of practical JSCC scheme for adaptive semantic rate control with IK-HARQ. 
Different from previous schemes,
we put forward a simplified system architecture with one encoder only for retransmission rather than multiple encoders.
Moreover, the end-to-end design is equipped with a self-adaptive semantic bit rate control mechanism to improve the transmission efficiency
and reduce the communication cost within different transmission environments. 
Besides, by introducing some denoising modules, the representation form of semantic information transmitted in the channel is effectively changed to further boost the denoising capability of semantic communications. 
The major contributions of this work are summarized as follows:

1) For the decoder side, we propose a novel JSCC scheme with semantic IK-HARQ for sentence transmission. Based on the recent  JSCC scheme \cite{[2]}, we use the transformer as semantic encoder and re-design the decoder to better leverage the information in the previously failed transmissions.
In particular, two different decoder schemes (i.e., a multi-decoder-based scheme and a single-decoder-based one) are proposed. By further utilizing the incrementally transmitted knowledge,
significant communication cost can be reduced on the basis of accurate semantic information transmission.

2) For the encoder side, in order to reduce the communication cost required to accurately convey semantic information, an end-to-end encoder mechanism with adaptive semantic bit rate control  is proposed. 
With a specific policy network, bit rate control is adaptively decided according to the semantic complexity of the content and the transmitted channel conditions,
which is shown to be more flexible than conventional rate control methods.

3) To further improve the reliability of  semantic transmission without using more bits,
we  adapt the basic deep neural network (DNN) structure of semantic coding (SC) and introduce a novel self-adaptive denoising module. 
Different from the SNR-adaptive denoising module which uses the SNR information of the channel for reducing the influence of channel noise and accurately conveys semantic information,
by incorporating self-attention mechanism into the self-adaptive denoising module, the self-adaptive denoising module only requires the context of semantic information instead of the accurate SNR information to realize self-denosing.

4) Through combining the above methods, we can realize adaptive semantic bit rate control with IK-HARQ at semantic transmission level.
The encoder with adaptive bit rate control enables transmission more robustly, while the decoder with IK-HARQ allows us to use the retransmitted sentences more effectively. The denosing module enhances the performance of the entire system in a complementary manner.

The remainder of this paper is organized as follows. Section II introduces the system model for semantic transmission.
The proposed end-to-end solution is explained  and discussed in Section III. In Section IV, we demonstrate the numerical results of the proposed schemes. Finally, Section V concludes this paper. 

\begin{figure*}[t]
	\centering
	\begin{minipage}{0.49\textwidth}
	\includegraphics[scale=0.265]{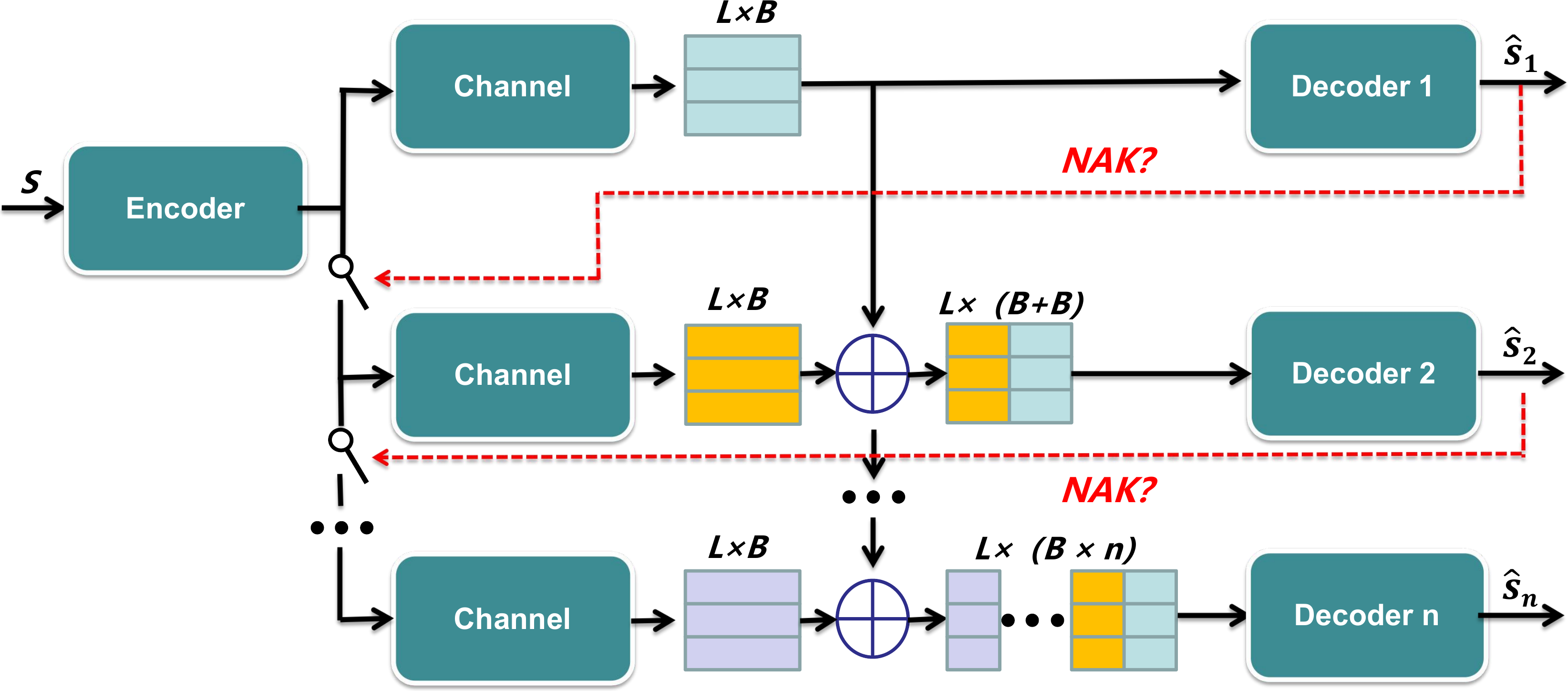}
	\caption{An illustration of the proposed semantic IK-HARQ with multi-decoders.}
	\end{minipage}
	\begin{minipage}{0.49\textwidth}
	\centering
	\includegraphics[scale=0.265]{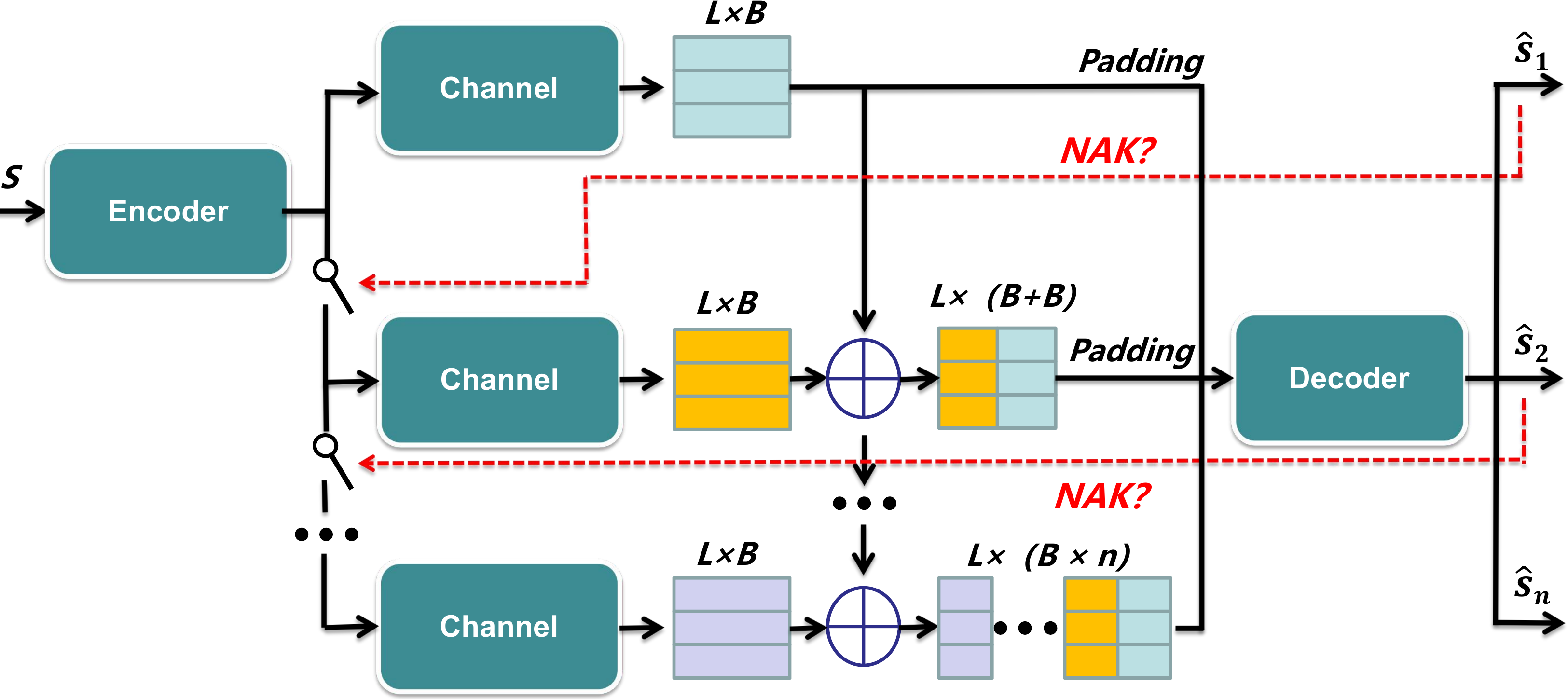}
	\caption{An illustration of the proposed semantic IK-HARQ with single decoder.}
	\end{minipage}
\end{figure*}

\section{System Model}

In this section, we will introduce the basic DL-based end-to-end transmission methods based on transformer for semantic transmission.


In traditional communication methods, to transmit a sentence \textbf{s}, the transmitter will convert it into bits by source coding and channel coding, which respectively plays the role of compressing data and reducing noise interference.
The process can be represented as 
\begin{equation}
\textbf{b} = \emph{C}_{\alpha}(\emph{S}_{\beta}(\textbf{s}))
\end{equation}
where $\emph{C}_{\alpha}(\cdot)$ and $\emph{S}_{\beta}(\cdot)$ represent the channel encoder through $\alpha$ algorithm and source encoder through $\beta$ algorithm, respectively.
After passing through the channel, $\hat{\textbf{b}}$ represents the received bits at the receiver. 
The received sentence is decoded as 
\begin{equation}
\hat{\textbf{s}} = \emph{S}^{-1}_{\beta}(\emph{C}^{-1}_{\alpha}(\hat{\textbf{b}}))
\end{equation}
where $\emph{S}^{-1}_{\beta}(\cdot)$ and $\emph{C}^{-1}_{\alpha}(\cdot)$ represent the source and channel decoding with the corresponding  $\alpha$ algorithm and $\beta$ algorithm.

In semantic communication, the traditional encoder $\emph{C}_{\alpha}(\emph{S}_{\beta}(\cdot))$
and decoder
$\emph{S}^{-1}_{\beta}(\emph{C}^{-1}_{\alpha}(\cdot))$  will be replaced by a DL-based encoder and decoder such as transformer or LSTM. 
In this paper,  we adopt transformer to act as the semantic encoder and decoder referred to DeepSC \cite{[3]} and the quantization process is referred to \cite{[6]}. The basic framework of semantic communications is shown in Fig. 1, which is used as the baseline mentioned in the follow-up experiment.
Specifically, the first step of semantic communications is to get the word embedding.
In that regard, we define the inputting sentences  \textbf{s} = [\emph{w}$_{1}$,\emph{w}$_{2}$,\dots,\emph{w}$_{\emph{L}}$], where  \emph{w}$_l$ ($l$$\in$ $\{1,\cdots,L\}$) represents the \emph{l}-th word in the sentences.
The embedding process can be written as
\begin{equation}
\textbf{X} = F_{\text{embed}}(\textbf{s};\gamma) = f_{\text{embed}}(\textbf{s};\gamma) + \textbf{PE}
\end{equation}
where \textbf{X} $\in$ $\mathbb{R}^{L \times K}$, \emph{L} is the length of the sentence, and \emph{K} is the symbol used for each word. 
$f_{\text{embed}}(\cdot;\cdot)$ represents the embedding layer, $\gamma$ is the parameter in embedding layer,  and the additive matrix \textbf{PE} is a constant matrix for position encoding  \cite{[13]}. 
After completing the embedding, the vector \textbf{X} will be fed into the semantic encoder.
The coding process firstly uses a semantic network (e.g. transformer) to  extract semantic features of input vectors
$\emph{F}_{\text{en}}(\cdot)$
and passes it through multilayer perceptron (MLP) $\emph{f}_{\text{en}}(\cdot)$ to better cope with the noise of channels.
For convenience, we denote the semantic coding process as
$\emph{SC}_{\text{en}}(\cdot;\alpha_{\text{en}})$, which can be written as
\begin{equation}
\textbf{s}_{\text{en}} =
\emph{SC}_{\text{en}}(\textbf{s};\alpha_{\text{en}}) = \emph{f}_{\text{en}}(\emph{F}_{\text{en}}(F_{\text{embed}}(\textbf{s})))
\end{equation}
where $\alpha_{\text{en}}$ represents all the trainable parameters in the encoder,  $\textbf{s}_{\text{en}}$ 
$\in$ $\mathbb{R}^{L \times M}$, and \emph{M} is the output dimension of MLP.
Completing the coding, the vector $\textbf{s}_{\text{en}}$ 
 will be converted to  $\mathbb{R}^{L \times B}$
according to one-bit quantization module,
where \emph{B} is the number of bits for each word. 
The quantization module is also consisted of MLP. 
We denote the quantization process as $Q(\cdot)$, and its output can be written as
\begin{equation}
\textbf{b} = 
\emph{Q}(\textbf{s}_{\text{en}};\theta_{\text{en}})
\end{equation}
Similar to the coding process, the decoding process can also be divided into two steps (i.e., dequantization and semantic decoding),
which can be denoted as $Q^{-1}(\cdot)$ and $\emph{SC}_{de}(\cdot)$.
$\emph{SC}_{\text{de}}(\cdot)$ is also consisted of an MLP     
 $\emph{f}_{\text{de}}(\cdot)$ and a semantic network $\emph{F}_{\text{de}}(\cdot)$. The decoding process can be denoted as 
\begin{equation}
\emph{SC}_{\text{de}}(\cdot) = 
F_{\text{de}}(\emph{f}_{\text{de}}(\cdot))
\end{equation}
Therefore the entire transmission process can be written as 
\begin{equation}
\hat{\textbf{s}} = 
\emph{SC}_{\text{de}}(\emph{Q}^{-1}(h(\emph{Q}(\emph{SC}_{\text{en}}(\textbf{s})))))
\end{equation}
where $h(\cdot)$ is the channel layer.

The entire training process of the semantic communication system can be divided into three steps.

1) Training the semantic coder (encoder \& decoder) without  quantization and dequantization layers. This training process can be denoted as 
\begin{equation}
(\alpha_{\text{en}},\alpha_{\text{de}})
=\arg \min \emph{L}_{\text{CE}}(\textbf{s},SC_{\text{de}}(SC_{\text{en}}(\textbf{s};\alpha_{\text{en}});\alpha_{\text{de}}))
\end{equation}
where $\emph{L}_{\text{CE}}(\cdot)$ is the cross-entropy (CE) loss function. 

2) Using the semantic encoder trained in the previous step to train the quantization module. This training process can be expressed as
\begin{equation}
(\theta_{\text{en}},\theta_{\text{de}})
=\arg \min \emph{L}_{\text{MSE}}(\textbf{s}_{\text{en}},Q^{-1}(Q(\textbf{s}_{\text{en}};\theta_{\text{en}});\theta_{\text{de}}))
\end{equation}
where $\textbf{s}_{\text{en}}$ = $SC_{\text{en}}(\textbf{s};\alpha_{\text{en}})$ is obtained by the semantic encoder, and the  loss function $\emph{L}_{\text{MSE}}(\cdot)$ is the mean-squared error (MSE).

3) Finally, finetune all trainable parameters,
\begin{equation}
(\theta_{\text{en}},\theta_{\text{de}},\alpha_{\text{en}},\alpha_{\text{de}})
=\arg \min \emph{L}_{\text{CE}}(\textbf{s},\hat{\textbf{s}}) 
\end{equation}

\section{Semantic Solution with  Adaptive Bit Rate Control and IK-HARQ}

\begin{algorithm}[t]
	\caption{The training method of the semantic IK-HARQ} 

	\label{alg:alg2}
 
        \textbf{Initialization:} The parameters of the decoder according to the preset number of retransmissions.
       
        \textbf{Input:} The transmitted sentence \textbf{s}.

        \textbf{Output:} The decoder $\emph{SC}_{\text{de}}(\cdot;\alpha_{\text{de}}),
        \emph{Q}^{-1}(\cdot;\theta_{\text{de}})$.
       
       \textbf{1):} \textbf{Transmitter:} $\textbf{b} = \emph{Q}(\emph{SC}_{\text{en}}(\textbf{s};\alpha_{\text{en}});\theta_{\text{en}})$.
       
       \textbf{2):} Randomly choose transmission times $i$, and transmit \textbf{b} $i$ times over the channel.
       
       \textbf{3):} \textbf{Receiver:}
       
       \quad \textbf{(1):} Zero padding  according to transmission times.
       
       \quad \textbf{(2):} $\hat{\textbf{s}_{i}} = \emph{SC}_{\text{de}}(\emph{Q}^{-1}([\hat{\textbf{b}}_{1},\cdots,\hat{\textbf{b}}_{i}];\theta_{\text{de}});\alpha_{\text{de}})$.
       
       \textbf{4):} Compute loss function $\emph{L}_{\text{CE}}$.
       
       \textbf{5):} Train $\theta_{\text{de}}, \alpha_{\text{de}}$ by gradient descent.
       

\end{algorithm}

In this section, we propose a semantic communication system with several novel semantic modules.
In the beginning, we introduce the semantic IK-HARQ, which uses only one encoder but employs a smarter decoder to better leverage the retransmitted messages by IK-HARQ.
Then,  we design the encoder with adaptive semantic bit rate control to improve the coding efficiency. 
Besides, through adding the denoiser to the coder, the representation form of the semantic vector is changed to reduce the influence of channel noise and accurately convey semantic information.
Finally, by integrating the above modules, we can get a more efficient ensemble solution with  adaptive bit rate control and IK-HARQ at the semantic level.

\subsection{Semantic IK-HARQ}

Although semantic JSCC \cite{[3]} has proven effective under low SNR or with the limited number of bits for each sentence, it still can not guarantee that the receiver can fully understand the meaning of the sender in every transmission.
Therefore, to avoid possible semantic misunderstanding encountered in semantic transmission in a lossy environment,  it is inevitable to combine semantic communications and HARQ.
In pioneering works, Jiang \emph{et al}. [25] initiate to combine HARQ and semantic communication. 
Jiang's study has confirmed the potential of combining HARQ with semantic communication. However, the presented  scheme involves multiple coders which may be cumbersome for the semantic communication system to practical use.
To further overcome these drawbacks, inspired by \cite{[7]}, which proposes to combine successive refinement with JSCC for image transmission, we devise a new  HARQ scheme by leveraging incremental knowledge (IK) that enables an efficient and more precise transmission whilst streamlining the practical design with one unified decoder.

\begin{algorithm}[t]
	\caption{Train the semantic communication system with adaptive semantic bit rate control } 

	\label{alg:alg2}
 
       
       \textbf{Step 1:} Train a multi-rate encoder.

        \textbf{Initialization:} The parameters of the decoder according to the preset bit length.
       
        \textbf{Input:} The transmitted sentence \textbf{s}.

        \textbf{Output:} $\emph{SC}_{\text{en}}(\cdot;\alpha_{\text{en}})$, $\emph{Q}(\cdot;\theta_{\text{en}})$, $\emph{SC}_{\text{de}}(\cdot;\alpha_{\text{de}}),
        \emph{Q}^{-1}(\cdot;\theta_{\text{de}})$.
        
       \textbf{1):} Follow the training scheme in Section II to get an encoder with the preset highest bit length.
       
       \textbf{2):} \textbf{Transmitter:} $\textbf{b} = \emph{Q}(\emph{SC}_{\text{en}}(\textbf{s};\alpha_{\text{en}});\theta_{\text{en}})$.
       
       \textbf{3):} Randomly choose bit length used for this transmission.
       
       \textbf{4):} After passing through channel, zero padding $\hat{\textbf{b}}$ according to chosen bit length.
       
       \textbf{5):} \textbf{Receiver:} $\hat{\textbf{s}} = \emph{SC}_{\text{de}}(\emph{Q}^{-1}(\hat{\textbf{b}};\theta_{\text{de}});\alpha_{\text{de}})$.
       
       \textbf{6):} Train $\theta_{\text{de}}, \alpha_{\text{de}}$, $\theta_{\text{en}}, \alpha_{\text{en}}$ by loss function $\emph{L}_{\text{CE}}(s,\hat{s})$.
       
       \textbf{Step 2:} Train policy network.
       
       \textbf{Initialization:} The parameters of the policy network.
       
        \textbf{Input:} The vector $\textbf{s}_{\text{en}}$.

        \textbf{Output:}The policy network $\theta_{\text{policy}}$.
       
       \textbf{1):} Relabel the training data set according to coder gotten in Step 1.
       
       \textbf{2):} Obtain the corresponding bit length label $\hat{l}$ by feeding   $\textbf{s}_{\text{en}}$ to the policy network.
       
       \textbf{3):} Train $\theta_{\text{policy}}$ by loss function $\emph{L}_{\text{CE}}(l,\hat{l})$.
       

\end{algorithm}

\subsubsection{Multi-Decoders}
\ 

The illustration of the proposed scheme is shown in 
Fig. 2. 
In this scheme, N different decoders are separately responsible for varied transmissions and collaborate with one shared encoder, namely
\begin{equation}
\textbf{b} = \emph{Q}(\emph{SC}_{\text{en}}(\textbf{s};\alpha_{\text{en}});\theta_{\text{en}})
\end{equation}
After receiving $\hat{\textbf{b}}_{i}$
from the \emph{i}-th transmission, 
the decoder concatenates it with previously transmitted bits and decodes from this concatenated information, yielding
\begin{equation}
\hat{\textbf{s}}_{i} = \emph{SC}_{\text{de}}(\emph{Q}^{-1}_{i}([\hat{\textbf{b}}_{1},\cdots,\hat{\textbf{b}}_{i}];\theta_{\text{de},i});\alpha_{\text{de},i})
\end{equation}
where $\hat{\textbf{s}}_{i}$ represents the received sentence after the \emph{i}-th transmission;
$\emph{Q}^{-1}_{i}$ ($i$ $>$ 1) adds a new integration module, which is consisted of MLP to the  original dequantization module; meanwhile $\emph{SC}_{\text{de}}$ has the same architecture as
in Fig. 1.
Notably, for this multi-decoder scheme, when \emph{i} = 1, we retain the same encoder and decoder as usual. Hence, there is no need for retraining  $\emph{SC}_{\text{en}}$, $\emph{SC}_{\text{de},1}$ and $\emph{Q}^{-1}_{1}$. 
As for the newly introduced  $\emph{SC}_{\text{de},i}$,
$\emph{Q}^{-1}_{i}$ ($i$ $>$ $1$), we train them recursively. 
In other words, for training of the \emph{i}-th decoder, we fix the parameters in the previous \emph{i}-1 decoder and train it as 
\begin{equation}
(\alpha_{\text{de},i},\theta_{\text{de},i})
=\arg \min \emph{L}_{\text{CE}}(\textbf{s},\hat{\textbf{s}_{i}}) 
\end{equation}
Finally, after the above training, we can get the semantic IK-HARQ with one encoder and multiple-decoders.
These extra trained decoders will be activated when the initially restored sentence can not pass the corresponding error detection standard (i.e., CRC or other semantic error detection codes). After receiving a negative acknowledgement signal (NAK), the transmitter will resend the encoded semantic information again, so that the \emph{i}-th decoder could utilize the nonlinear mapping to aggregate IK that is transferred  in the previous \emph{i}-1 times. Therefore,  the performance of transmission will be progressively improved with the accumulation of IK.


\subsubsection{Unified Single Decoder}
\

Different from the previous scheme with multi-decoders,
we can alternatively use only a single decoder for IK-HARQ, as described in Fig. 3.
Same as before, the quantized bit vector \textbf{b} is used for transmission.
Since only one decoder is used for retransmission, the decoder has to be trained for different code sizes to retrieve information from the codeword part by part, 
and if the transmission times do not reach the preset maximum transmission times,
those extra dimensions are automatically padded with zero vectors.
Other parts of the scheme are identical to the multi-decoder solution. However, the training process could be significantly different.
As shown in Algorithm 1,
during the training, we will randomly select the transmission times required for this transmission.
If the chosen transmission times do not reach the preset maximum number of retransmissions, the retransmitted bits with zero padding will be fed into the decoder.
In this way, the system could learn to extract effective semantic information from 
the retransmitted  semantic knowledge.



\begin{figure*}[htbp]
	\centering
	\includegraphics[scale=0.33]{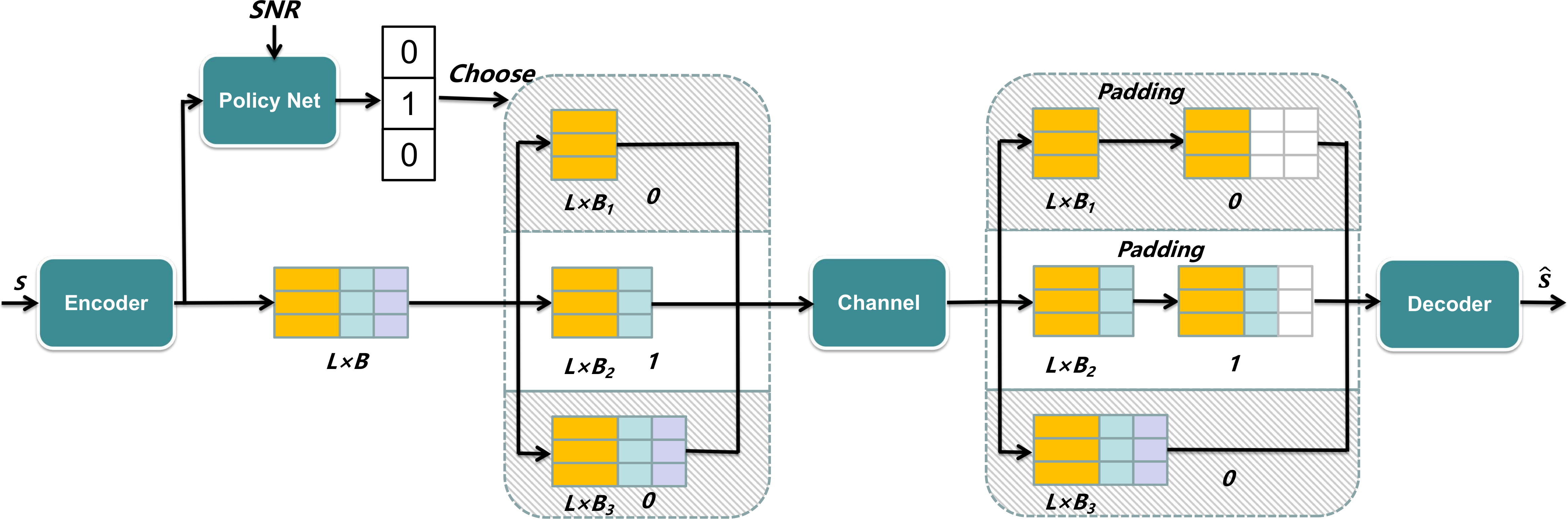}
	\caption{An illustration of SC with adaptive semantic bit rate control.}
\end{figure*}

\begin{figure}[t]
	\centering
	\includegraphics[scale=0.29]{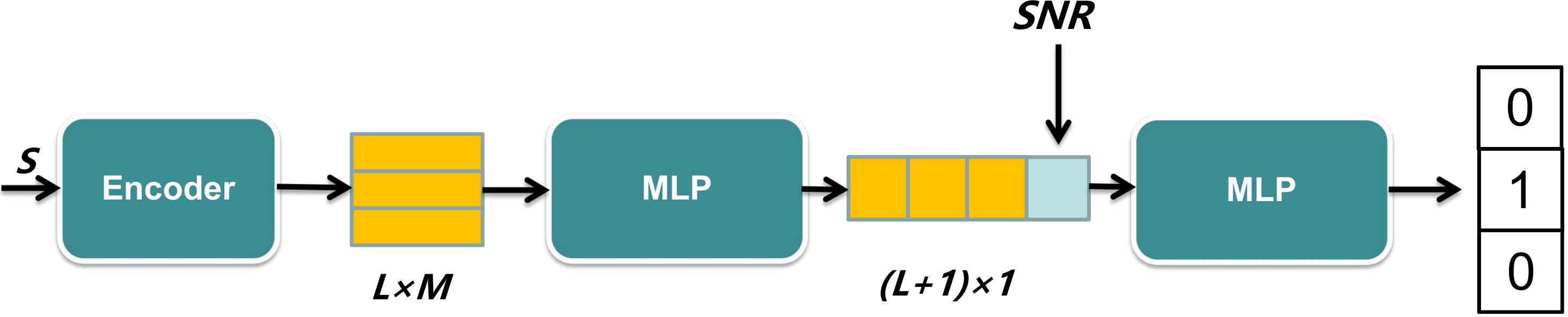}
	\caption{The neural network structure of policy network.}
\end{figure}

\subsection{Adaptive Semantic Bit Rate Control}

Aside from the above mentioned IK-HARQ scheme which mainly concentrates on the decoder-side accuracy and efficiency design, in this part we introduce another technique that further allows an adaptive source-level bit-rate control. Through comprehensively considering semantics and SNR, we hope the encoder can transmit with appropriate bit length at the beginning to reduce transmission error and save bit resources as much as possible.
Specifically, we divide the process of designing a semantic communication system with adaptive semantic bit  rate control  into two steps.
The first step is to design an encoder that is capable of using different bit lengths to transmit sentence. 
Through which, the system supports transmission at different bit lengths.
The second step is to design a specific policy network to choose the suitable bit length for the multi-rate encoder. 
In this vein, we can flexibly adjust the bit length based on different conditions of each transmission.
The complete framework of this scheme can be seen in  Fig. 4.
For simplicity of representation, we assume there are three bit lengths to choose.

Actually, relying on the robustness of the semantic system, even if we only transmit part of bits correctly, the receiver can also decode the correct semantics information and the length of the used bitstream will only affect the accuracy of transmission.
This phenomenon  can be commonly observed in the literature (i.e. \cite{[6],[3]}), 
however, existing works are mainly designed at a fix bit-length, which hampers the flexibility. According to this phenomenon, we can design a multi-rate encoder through transmitting segmentation of bits to meet the requirements of different channels and save transmission resources.
In this transmission process, through the semantic encoder $\emph{SC}_{\text{en}}(\cdot;\alpha_{\text{en}})$ 
and quantification model $\emph{Q}(\cdot;\theta_{\text{en}})$,
the transmitted sentence \textbf{s} will be converted to the bitstream \textbf{b} with length  $L \times B$, where \emph{L} is the length of the sentence and $B$ represents the bit length which can be further subdivided into $B_{1}$, $B_{2}$, $\cdots$, $B_{n}$ ($B_{1}$ $<$ $B_{2}$ $<$ $\cdots$ $<$ $B_{n}$ $<$ $B$).
In this scheme, the selected bit length  is contingent on the channel quality and the semantic complexity of the messages. More specifically, larger bit length is selected for poorer channels and more complex semantic messages. 
After choosing  a different bit length for transmission, the decoder will automatically pad the input bit vectors $\hat{\textbf{b}}$ 
to a fixed dimension $L\times B$.
Then the subsequent decoding is the same as that in the previous scheme
in Section II. 
The detailed training process is shown in Algorithm II.
First of all, the encoder with the preset  highest
bit length is trained as the same process in Section II. Afterwards, by randomly selecting bit length from those preset ones,
the encoder could learn to transmit semantic information at different bit lengths.

In addition to the multi-rate encoder, the policy network is also indispensable in order to realize adaptive bit rate control.
The unique DNN structure of the policy network  is presented in Fig. 5.
Different from the previous bit control schemes which usually ignore semantics, considering the particularity of semantic communication, the policy network will comprehensively  take account of the semantic complexity of the sentence and the transmission channel SNR, so as to make rate decision,
which can be written as
\begin{equation}
\hat{l} =  f_{2} (\text{Concatenate}(f_{1}(\textbf{s}_{\text{en}}), \text{SNR}))
\end{equation}
where $\hat{l}$ is the selected bit length label, $f_{1}$ and $f_{2}$ are the corresponding leranable non-linear mappings used in the policy network. The  semantic vector $\textbf{s}_{\text{en}}$ can be obtained by (4),
while SNR measures the quality of a transmission channel.

Next, we discuss how to train the policy network in practice,
for which the detailed process is also shown in Algorithm II.
With the help of the aforementioned multi-rate encoder, it becomes possible to testify the performance of different sentences under different SNR situations at the selected bit length, and constitute these results into a new training dataset.
Afterwards, based on this dataset, it is viable to get the minimum bit length on the premise of ensuring correct semantic transmission, thereby transforming the aforementioned dataset into a more comprehensive one with 
the optimal bit length under the corresponding channel SNR.
Now, it becomes ready to train the policy network with 
the sentence and channel SNR as the input of the policy network.
The corresponding optimal bit length can be obtained by optimizing the loss function
\begin{equation}
\theta_{\text{policy}}
=\arg \min  \, \emph{L}_{\text{CE}}(l,\hat{l}) 
\end{equation}
where $\theta_{\text{policy}}$ is the parameters of the policy network, 
\emph{l} denotes the   optimal length label that we have tested with the  coder before; $\hat{l}$ is the length label that is deduced by the policy network.

\begin{figure}[t]
	\centering
	\includegraphics[scale=0.31]{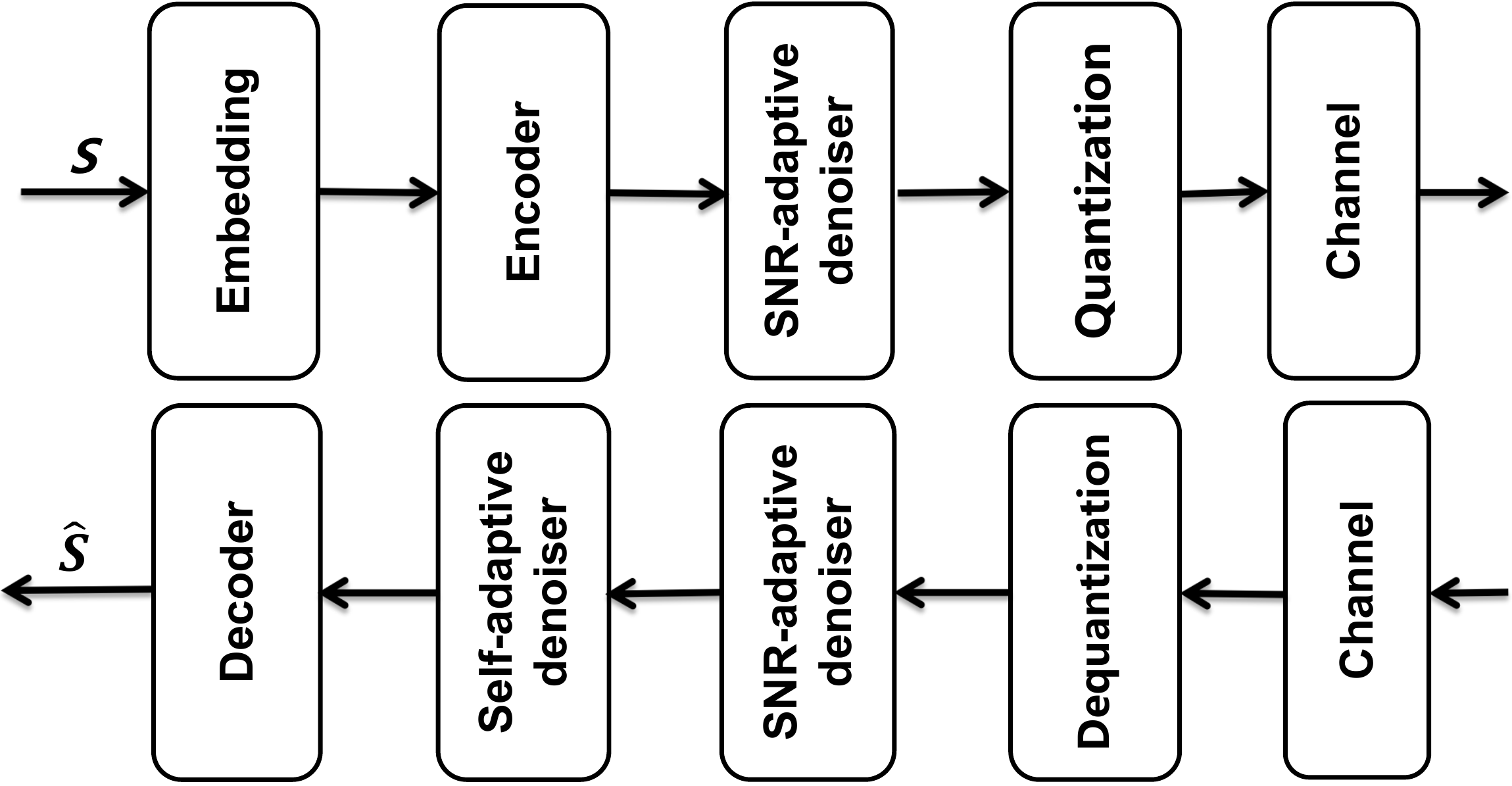}
	\caption{A general illustration of the proposed SC with denoising modules.}
\end{figure}

\begin{figure}[t]
	\centering
	\includegraphics[scale=0.275]{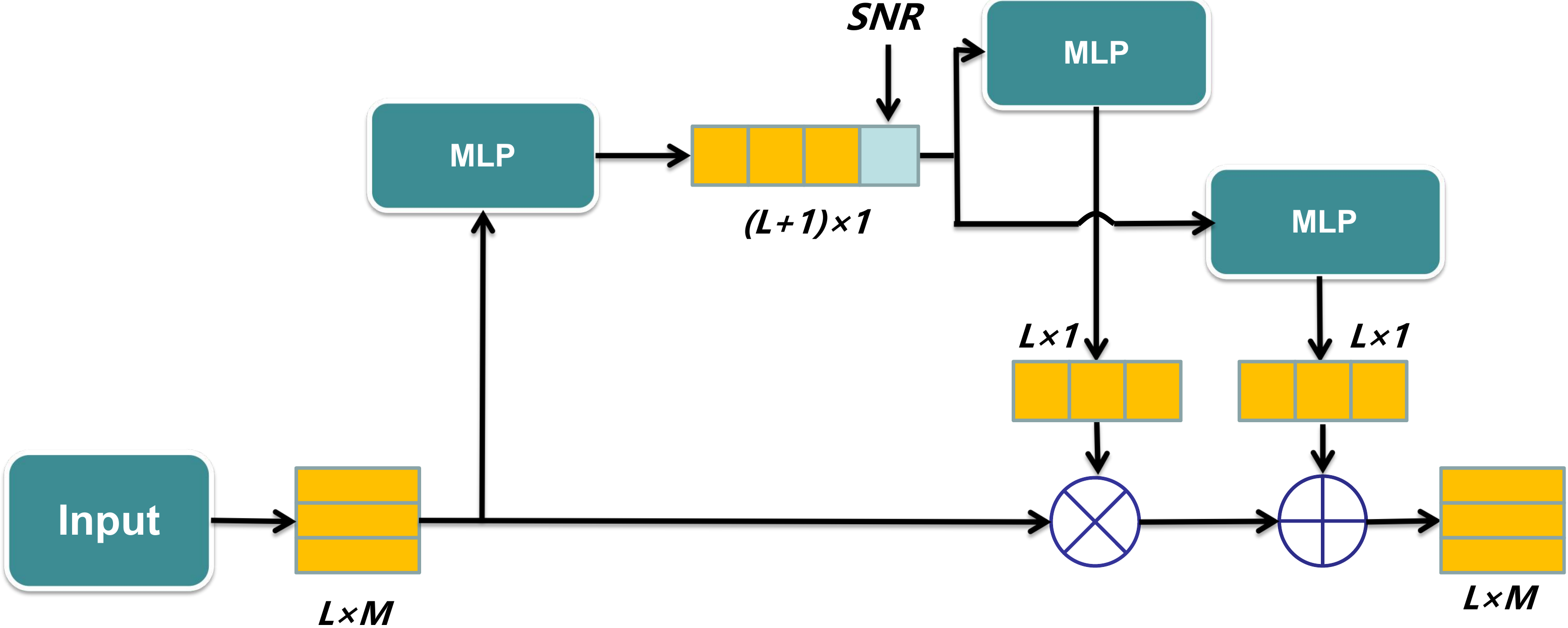}
	\caption{The DNN structure of SNR-adaptive denoising module.}
\end{figure}

\begin{figure}[t]
	\centering
	\includegraphics[scale=0.29]{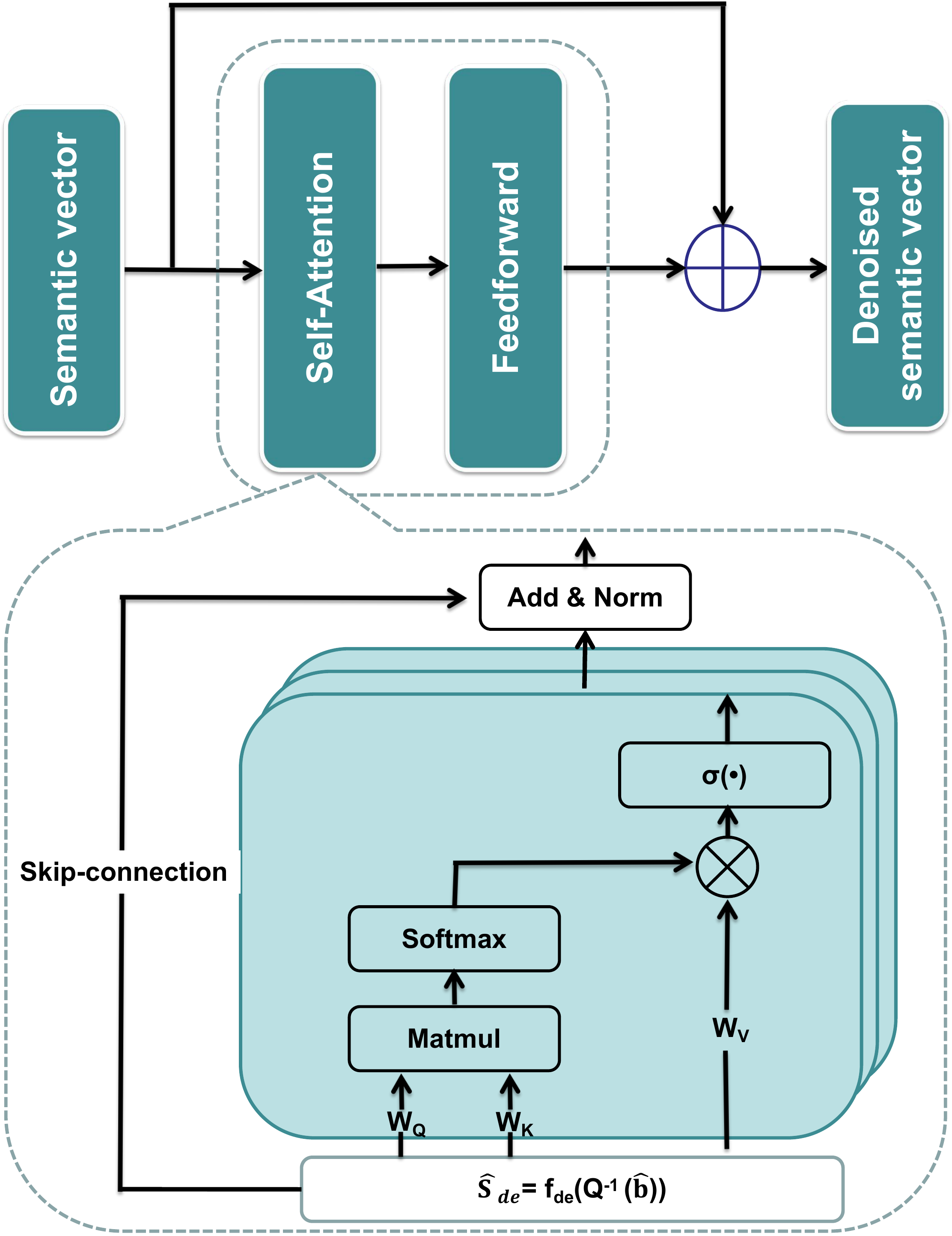}
	\caption{The DNN structure of self-adaptive denoising module.}
\end{figure}

\subsection{Self-adaptive Denoising Module}

Although the  two methods above can significantly improve the coding efficiency of communications,
these methods inevitably use more bits in exchange for performance improvement.
Therefore, in order to fully exploit the potential of semantic communications,
we consider a novel method to further boost the performance of the system without using more bit resources.

The complete framework of the proposed semantic communication system is shown in  Fig. 6.
In particular, we leverage
the SNR-adaptive denoising module as well as the self-adaptive denoising module 
to better enhance the semantic robustness.

Here, the SNR-adaptive denoising module is mainly referred to \cite{[8],[36]} and the MLP-based structure of the SNR-adaptive denoising module is shown in 
Fig. 7. 
The input semantic vector $\textbf{s}_{\text{en}}$ $\in$ $\mathbb{R}^{L \times M}$, which can be obtained by (4), 
is first pooled to an $L$-length vector 
and then concatenated with the SNR value.
Afterwards, the concatenated vector is passed through two different nonlinear mappings to generate the factors for channel-wise scaling
and addition.
Therefore, after exploiting these factors, the SNR-adaptive model can realize its denoising function.
Typically, two SNR-adaptive denoising modules, which share the same DNN structure but contain different parameters, can be utilized in the encoder and decoder.
The first module plays the role of precoding to strengthen the denoise ability of the semantic vector; 
while the second module eliminates the noise in the semantic vector based on the channel SNR.

The DNN structure of the self-adaptive denoising module is shown in
Fig. 8. This module is composed of a multihead self-attention layer and adopts the structure of ResNet \cite{[16]}.
This module is independent of the SNR-adaptive module and can be used alone without the SNR information apriori.
The input of the module is the semantic vector $\hat{\textbf{s}}_{\text{de}}$, which 
can be obtained from the bits $\hat{\textbf{b}}$, by passing through the dequantization module and the MLP, yielding
\begin{equation}
\hat{\textbf{s}}_{\text{de}}
=f_{\text{de}}(Q^{-1}(\hat{\textbf{b}}))
\end{equation}
In other words, after dequantization, a semantic vector is restored from a simple binary vector, thus laying the unique foundation  to use self-attention mechanism.
Despite suffering from the channel noise interference, $\hat{\textbf{s}}_{\text{de}}$ can not be dequantitated as same as ${\textbf{s}}_{\text{en}}$, there still remains some useful semantic information between the contexts. 
Therefore, with the help of attention
mechanism to assign different weights to distinguish the importance of the input, we can obtain
\begin{equation}
\text{Attention}(\hat{\textbf{s}}_{\text{de}}) = \text{Softmax}(QK^{T})V 
\end{equation}
where  $Q = W_{Q}$  $\cdot$     
$\hat{\textbf{s}}_{\text{de}}$, 
$K = W_{K}$  $\cdot$  
$\hat{\textbf{s}}_{\text{de}}$, 
$V= W_{V}$  $\cdot$
$\hat{\textbf{s}}_{\text{de}}$, 
are the results of affine transformations determined by the weights $W_{Q}$, $W_{K}$, $W_{V}$ respectively,
and the superscript \emph{T} denotes the matrix transpose operation. 
Consequently, the corresponding importance weights can be obtained by computing  $\text{Softmax}(QK^{T})$, and $V$ is the value needed to be scaled. 
Benefiting from the self-attention mechanism, we can get the corresponding importance weights about $\hat{\textbf{s}}_{\text{de}}$, so that it becomes possible to resolve cases suffering semantic interference and wisely scale the information to avoid it. The representation form of a semantic vector will be changed to highlight semantics and reduce the semantic error after scaling.
In this way, the self-adaptive denoising module can help the semantic vector denoise itself without using  the SNR information.

\newcommand{\tabincell}[2]{\begin{tabular}{@{}#1@{}}#2\end{tabular}}  

\begin{table*}[htb]
  \centering
  \caption{THE SETTING OF SEMANTIC COMMUNICATION SYSTEM}
\setlength{\tabcolsep}{5mm}{
\begin{threeparttable}
\begin{tabular}{c|c|c|c|c}
\hline
{Transmission System} & {Modules} & {Layer} & \tabincell{c}{Output\\dimensions} & \tabincell{c}{Activation\\function}\\ \hline
\multirow{11}{*}{SC} & Input  & \textbf{s}  & 30 & / \\ \cline{2-5} 
 & {$\text{SC}_{\text{en}}$}  & \tabincell{c}{Embedding\\3$\times$ Transformer\\MLP}  & \tabincell{c}{(30,128)\\(30,128)\\(30,16)} & \tabincell{c}{/ \\ /\\ReLU} \\ \cline{2-5} 
 & $Q$ & \tabincell{c}{MLP\\ 1-bit quantization} &  \tabincell{c}{(30,30)\\(30,30)} &\tabincell{c}{ReLU\\/} \\ \cline{2-5}  & $Q^{-1}$ & MLP &  (30,16) &ReLU \\ \cline{2-5}
 & {$\text{SC}_{\text{de}}$}  & \tabincell{c}{MLP\\3$\times$ Transformer\\MLP}  & \tabincell{c}{(30,128)\\(30,128)\\(30,32478)} & \tabincell{c}{ReLU \\ ReLU\\SoftMax} \\  \cline{2-5}
 & Output  & $\hat{\textbf{s}}$  & 30 & / \\ \cline{2-5} \hline
Channel & AWGN & /& /& / \\ \hline
\multirow{7}{*}{\tabincell{c}{IK-HARQ\\(Multi-Decoders)}} & Input  & \textbf{s}  & 30 & / \\ \cline{2-5} 
 & Encoder & \tabincell{c}{$\text{SC}_{\text{en}}$\\$Q$} & \tabincell{c}{(30,16)\\(30,30)} &\tabincell{c}{/\\ReLU}  \\ \cline{2-5}  
 & Decoder $i$ & \tabincell{c}{Input\\$Q^{-1}$\\$\text{SC}_{\text{de}}$} & \tabincell{c}{(30,$30\times i$)\\(30,16)\\(30,32478)} &\tabincell{c}{/\\ReLU\\SoftMax}  \\ \cline{2-5}  
 & Output  & {$\hat{\textbf{s}}$}  & 30 & / \\  \hline
\multirow{7}{*}{\tabincell{c}{IK-HARQ\\(Single Decoder)}} & Input  & \textbf{s}  & 30 & / \\ \cline{2-5} 
 & Encoder & \tabincell{c}{$\text{SC}_{\text{en}}$\\$Q$} & \tabincell{c}{(30,16)\\(30,30)} &\tabincell{c}{/\\ReLU}  \\ \cline{2-5}  
 & \tabincell{c}{Decoder \\($i$-th transmission)} & \tabincell{c}{Padding\\$Q^{-1}$\\$\text{SC}_{\text{de}}$} & \tabincell{c}{(30,$30\times i$)\\(30,16)\\(30,32478)} &\tabincell{c}{/\\ReLU\\SoftMax}  \\ \cline{2-5}  
 & Output  & {$\hat{\textbf{s}}$}  & 30 & / \\  \hline
 \multirow{7}{*}{\tabincell{c}{Multi-Rate\\ Encoder}} & Input  & \textbf{s}  & 30 & / \\ \cline{2-5} 
 & Encoder & \tabincell{c}{$\text{SC}_{\text{en}}$\\$Q$} & \tabincell{c}{(30,16)\\(30,60)} &\tabincell{c}{/\\ReLU}  \\ \cline{2-5}  
 & \tabincell{c}{Decoder } & \tabincell{c}{Padding\\$Q^{-1}$\\$\text{SC}_{\text{de}}$} & \tabincell{c}{(30,60)\\(30,16)\\(30,32478)} &\tabincell{c}{/\\ReLU\\SoftMax}  \\ \cline{2-5}  
 & Output  & {$\hat{\textbf{s}}$}  & 30 & / \\  \hline
 \multirow{4}{*}{Policy Network} & Input  & Vector  & (30,16) & / \\ \cline{2-5} 
 & MLP & \tabincell{c}{MLP\\Concatenating\\MLP} & \tabincell{c}{(30,1)\\(31,1)\\3} &\tabincell{c}{ReLU\\/\\ReLU}  \\  \hline
  \multirow{5}{*}{\tabincell{c}{SNR-adaptive\\Denoiser}} & Input  & Vector  & (30,16) & / \\ \cline{2-5}  
 & Denoiser & \tabincell{c}{MLP\\Concatenating\\MLP\\MLP} & \tabincell{c}{(30,1)\\(31,1)\\(30,1)\\(30,1)} &\tabincell{c}{ReLU\\/\\ReLU\\ReLU}  \\  \hline
  \multirow{3}{*}{\tabincell{c}{Self-adaptive\\Denoiser}} & Input  & Vector  & (30,128) & / \\ \cline{2-5} 
 & Denoiser & \tabincell{c}{Multihead\\self-attention} & (30,128) & /  \\  \hline
\end{tabular}
\begin{tablenotes}
\item[1] For convenience, $\text{SC}_{\text{en}}$ is directly used in the layer column to represent the same layers as shown in the $\text{SC}_{\text{en}}$ module of the SC. This applys to $\text{SC}_{\text{de}}$ and $Q$ as well.
\item[2] Decoder $i$ represents the $i$-th Decoder in IK-HARQ (Multi-Decoders), and Decoder ($i$-th transmission) means the preset maximum retransmission number is $i$.
\end{tablenotes}
\end{threeparttable}}
\end{table*}


\subsection{The Integrated End-to-End Solution}

\begin{figure*}[htbp]
	\centering
	\includegraphics[scale=0.32]{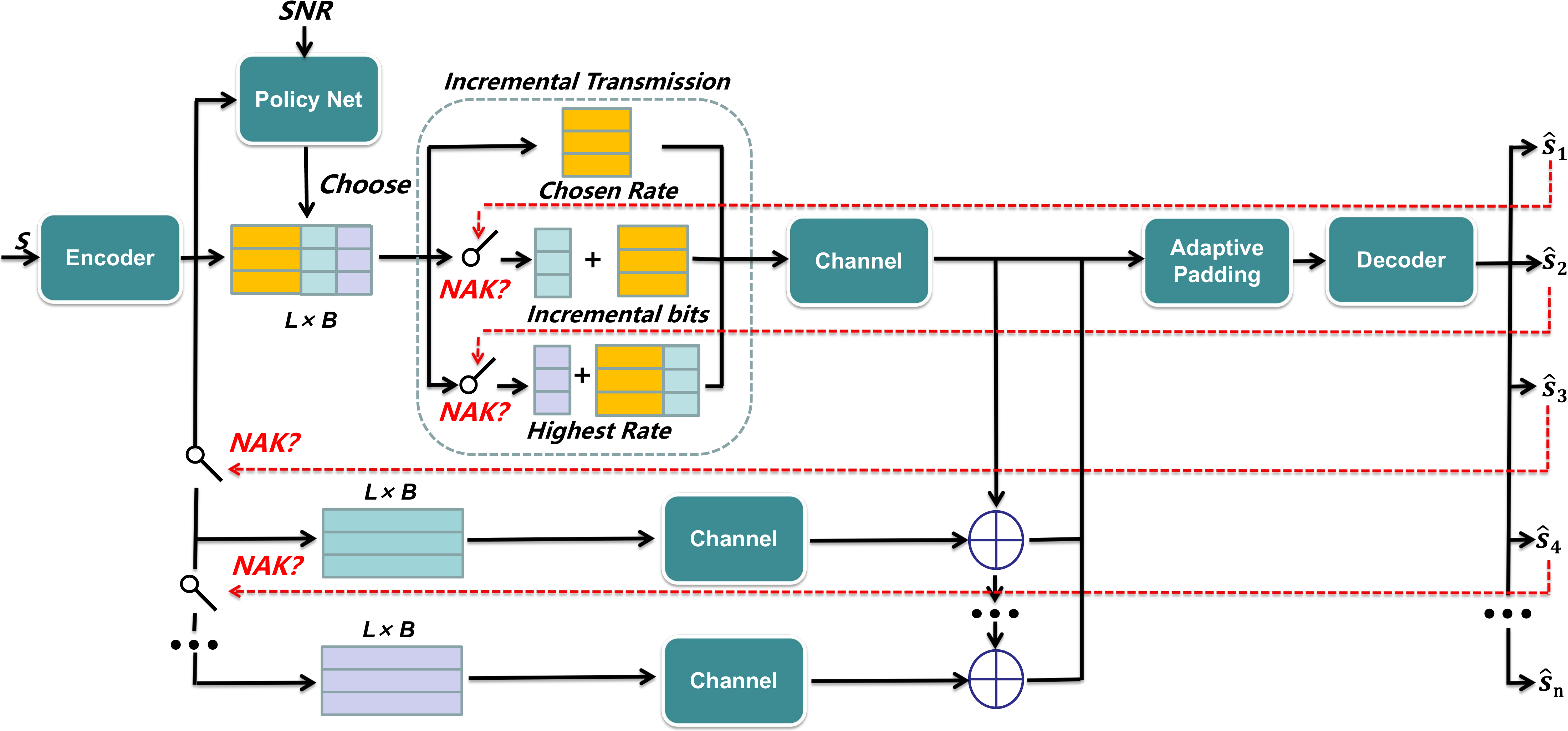}
	\caption{An illustration of the whole integrated end-to-end JSCC with semantic bit rate control.}
\end{figure*}

Finally, we integrate the three schemes proposed above and establish an ensemble  JSCC scheme with adaptive semantic bit rate control and IK-HARQ. As illustrated in Fig. 9,
this JSCC scheme makes considerable modifcations from the transmitter to the receiver  compared with the basic one described in Fig. 1.
First of all, for obtaining the entire semantic communication system, we need to test the transmission performance under different bit lengths in advance and choose several suitable bit lengths $B_{1}$, $B_{2}$ $\cdots$ $B_{n}$, $B$ ($B_{1}$ $<$ $B_{2}$ $<$ $\cdots$ $<$ $B_{n}$ $<$ $B$) to handle different transmission situations.
Once the bit length has been decided, we need to retrain the entire semantic system if we want to change the chosen bit lengths.
Specifically,
we can get the semantic encoder with adaptive bit rate control as in Section III-B.
Then we turn our attention to the receiver.
For the decoder, it is also important for us to decide the preset maximum transmission times. Similar to the encoder, since the DNN structure of the decoder depends on the maximal transmission times, its modification also requires a retraining.
The two decoder schemes (i.e., multi-decoders and single decoder) in Section III-A are appealing candidates. Here, the right part of Fig. 9 presents the outcome if we adopt the unified single decoder as in Fig. 3.
For the denoiser mentioned in Section III-C, we can optionally choose and integrate it into the semantic system. 

To transmit a sentence \textbf{s}, firstly, the policy network will choose the most suitable bit length $B_{i}$ according to the sentence and channel SNR.
After passing through the channel, the receiver will pad the received bits 
$\hat{\textbf{b}}$, and restore them to sentence $\hat{\textbf{s}}$.
However, due to channel uncertainty, the policy network can not  fully guarantee the accuracy of transmission.
Therefore when meet semantic errors, we need to use an NAK to inform the transmitter.
Here, besides adopting IK-HARQ, incremental transmission (IT) is also introduced 
as shown in Fig. 9, while being similar to traditional methods, we will gradually transmit the remaining bits (e.g. $L$ $\times$ ($B_{i+1}-B_{i}$), $\cdots$, $L$ $\times$ ($B_{n}-B_{n-1}$), $L$ $\times$  $(B-B_{n})$) to increase the bit length for improving the  transmission performance.
If the last remaining bits $L$ $\times$ ($B-B_{n}$) have been transmitted but the transmitter still receives a retransmitting signal, the transmitter will retransmit the entire bit vector. In this case, the decoder concatenates it with the previously transmitted bits to get useful semantic information. This process will be repeated until the successful transmission or reach the preset maximum transmission times.
What's more, if we can not give the SNR of the transmitted channel or  the difference between the estimated SNR and the actual SNR is too large, \emph{IT} can also help to save bit resources while ensuring transmission performance without using policy net.

The training methods of each module have been detailed above. Here we mainly summarize the entire training processes:

\begin{itemize}

\item[(1)] Set up the whole system according to the maximum bit length and transmission times.

\item[(2)] Train the whole neural network for IK-HARQ by randomly choosing  transmission times, as in Section III-A.

\item[(3)] Train the whole neural network with multiple lengths by randomly masking, as in Section III-B.

\item[(4)] After integrating the corresponding denoising module, 
retrain the entire semantic communication system, as in Section III-C.

\item[(5)] Use the obtained semantic communication system to update the dataset for the training of the policy network.

\end{itemize}

Notably, the aforementioned modules are mutually independent and can be used optionally.
Therefore, we can achieve an end-to-end semantic solution with adaptive semantic bit rate control and IK-HARQ.

\begin{figure*}[htbp]
\begin{minipage}{0.49\textwidth}
\centering
\includegraphics[scale=0.33]{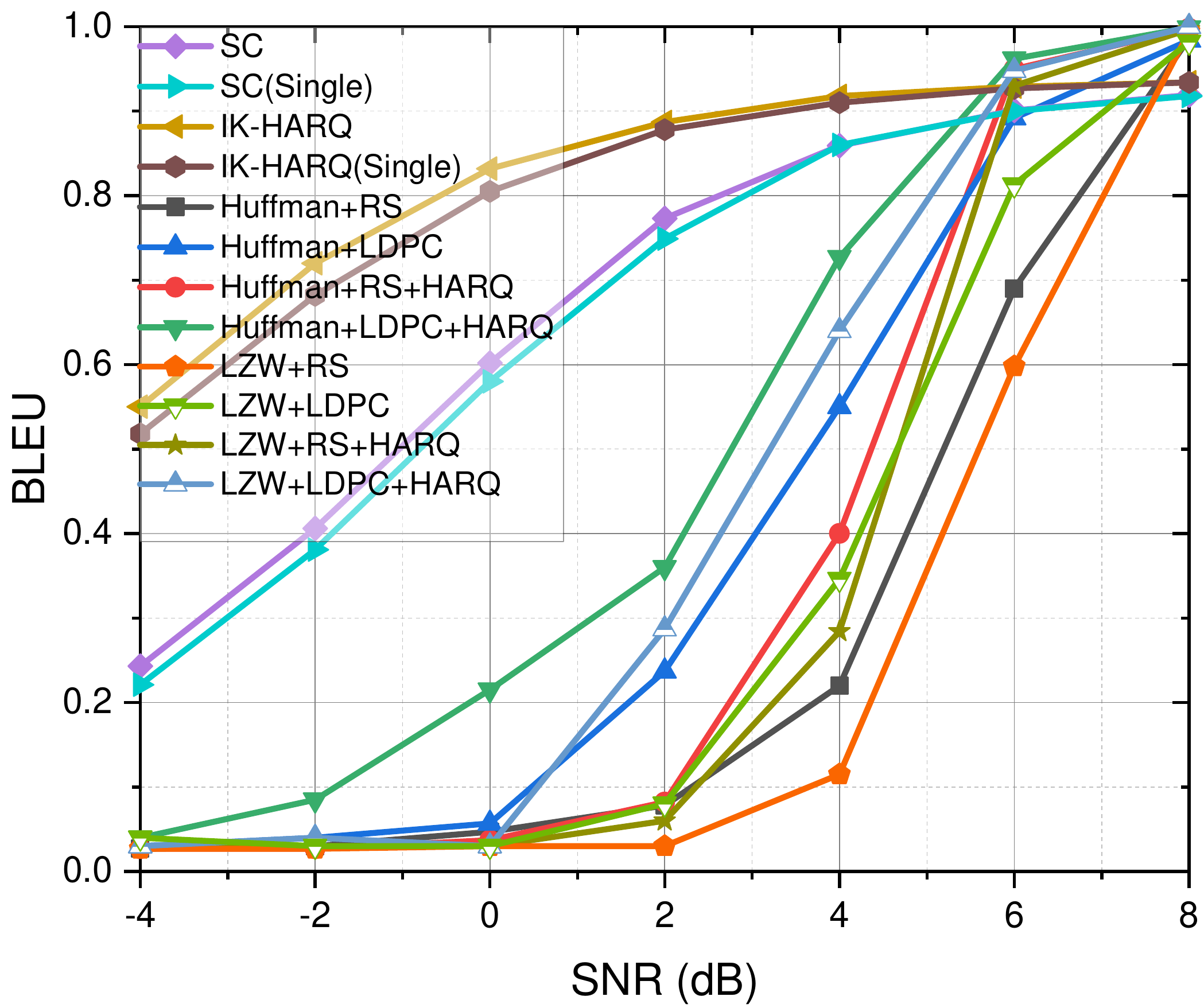}
\caption{BLEU score versus SNR for IK-HARQ with multiple or single decoder.}
\end{minipage}
\begin{minipage}{0.49\textwidth}
\centering
\includegraphics[scale=0.33]{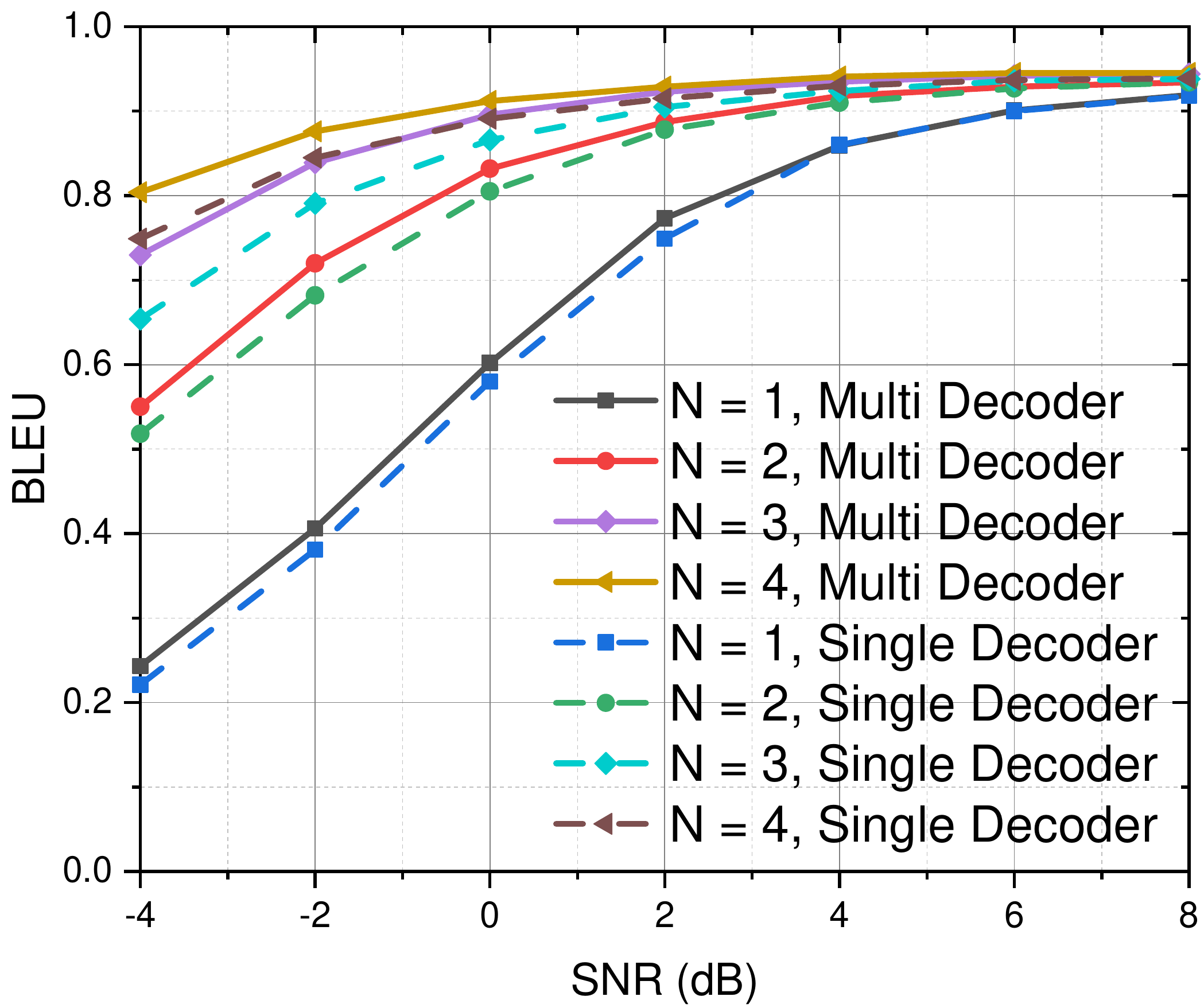}
\caption{BLEU score versus SNR for IK-HARQ with different transmission times.}
\end{minipage}
\end{figure*}

\begin{table}[t]
  \centering
  \caption{Average bits for each sentence}
\begin{threeparttable}
\begin{tabular}{c|c|c|c}
\hline
{Coding Scheme} & {B} & {Bit rate} & Bits/sentence\\ \hline
SC & 30  & /  & 513 \\ \hline
Huffman & /  & /  & 421 \\ \hline
Huffman + RS & /  & 5/7  & 589 \\ \hline
Huffman + LDPC & /  & 5/7  & 589 \\ \hline
Huffman + RS + HARQ & /  & 5/12  & 1010 \\ \hline
Huffman + LDPC + HARQ & /  & 5/12  & 1010 \\ \hline
LZW & /  & /  & 467 \\ \hline
LZW + RS & /  & 5/7  & 653 \\ \hline
LZW + LDPC & /  & 5/7  & 653 \\ \hline
LZW + RS + HARQ & /  & 5/12  & 1121 \\ \hline
LZW + LDPC + HARQ & /  & 5/12  & 1121 \\ \hline
\end{tabular}
\end{threeparttable}
\end{table}

\section{Simulation Settings and Numerical Results}
In this section, we will present the numerical simulation results of our introduced frameworks and discuss the
pros and cons of these semantic communicaton mechanisms.   
We will also compare their bit consumption with
traditional ones.
The implementation cost and computational complicity are also provided for completeness.

\subsection{Simulation Settings}
The adopted linguistic dataset is the standard proceedings of the European Parliament \cite{[17]}, which consists of around 2.0 million sentences and 53 million words. 
The vocabulary size in this experiment is 32478.
The dataset is pre-processed into lengths of sentences with 4 to 30 words.  

We present the detailed settings of the proposed semantic communication system in Table I.
The wireless channel used in this experiment is additive white Gaussian noise (AWGN).
The training SNR for all semantic communication systems mentioned above is in the range of -2dB to 6dB, which means that during the training, each transmission will  randomly choose  a channel SNR between  -2dB -- 6dB.
We use the JSCC scheme mentioned above in Section II as the baseline. 
Conventional source coding and channel coding schemes such as Huffman coding \cite{[38]}
and Lempel-Ziv-Welch (LZW) \cite{[37]} for source coding, and Low Density Parity Check Code (LDPC)  \cite{[18]} or RS coding \cite{[19]} for channel coding  are also employed in this experiment. 
In particular, the average length of coded bits are compared with the SC in Table II,   
the code rates of each transmission are 5/7 and 5/12,
so that we can compare the performance of various coding schemes under  approximately
the same bit rate.
Besides, for all experiments, we implement CRC where an NAK signal  will be sent to the transmitter when the estimated sentences at the receiver are detected to be incorrect.
Furthermore, the widely adopted evaluation metric in NLP, namely  bilingual evaluation understudy (BLEU) metrics \cite{[20]} is used to measure the performance.

\subsection{Numerical Results}

Fig. 10 shows the BLEU performance of semantic IK-HARQ with multiple or single decoder competing with the conventional methods. 
In this figure, 
for traditional coding schemes, we first observe that LDPC outperforms RS in both LZW and Huffman source coding.
When SNR is greater than or equal to 6 dB, LDPC coding can achieve almost 100$\%$ word accuracy, but RS coding performs much worse than LDPC. 
Moreover,  SC based approaches achieve  better results at the same SNR but require less transmission bits (approximately 513 bits) compared with the traditional ones (589 bits for Huffman or 653 bits for LZW).
Typically under low SNR levels, semantic communication can give full play to its advantages.
After retransmission, the performance of BLEU is significantly improved by increasing the code rate of channel coding.
However, different from the traditional ones using additional error correction bits, 
through retransmitting the semantic encoding bits, IK-HARQ with multi-decoders integrates the effective parts of two sequential transmissions and can also achieve a good result.
Also it can be  observed from  Fig. 10 that even with retransmission, the traditional coding methods can not work properly when SNR is equal or worse than 0 dB, 
while the semantic-based SC is still effective with the help of HARQ. Therefore, it can safely come to a conclusion that the general meaning of sentences has been successfully interpreted in the receiver.
What's more, IK-HARQ with single decoder show its superiority  in both the first transmission (\emph{SC (single)}) and retransmission
(\emph{IK-HARQ (single)}), 
especially when SNR is lower such that the traditional coding method can not work at a roughly similar bit length.
However, due to using only a single decoder for HARQ, there still exist some differences. The BLEU performance with a single decoder is inferior to the one with multi-decoders. 
However, it saves some space resources and training time at the cost of sightly lower accuracy, which we think is acceptable.
Although the SC performs better when SNR is less than 6 dB, SC can not  guarantee  100 \% accuracy compared with the traditional coding methods when SNR increases to  8 dB.
In other words, the semantic coding may still commit mistakes under tiny transmission errors or even no transmission errors,
which needs to be investigated further.

\begin{figure}[t]
	\includegraphics[scale=0.35]{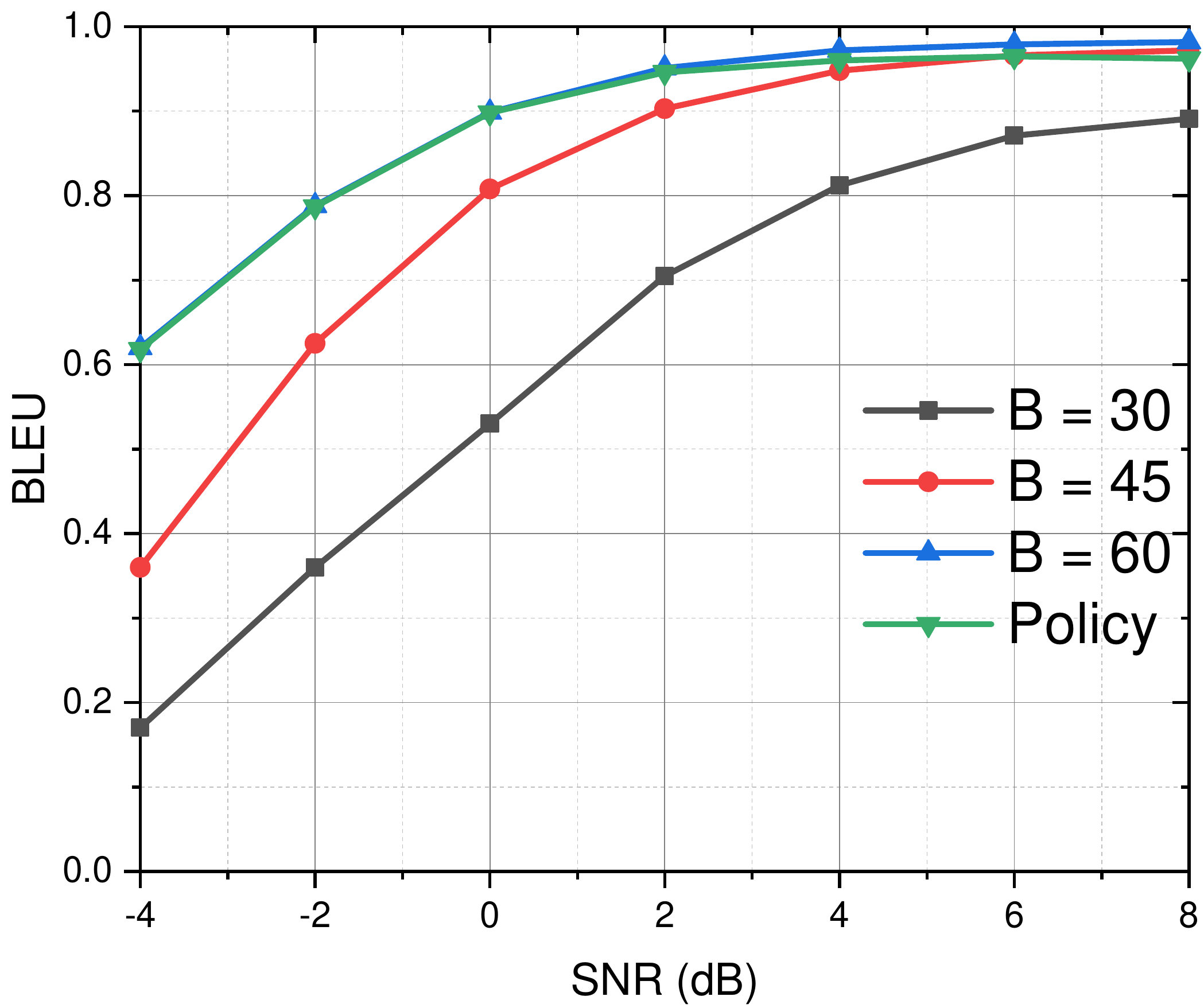}
	\caption{BLEU score versus SNR for SC with and without adaptive bit rate control.}
\end{figure}

On top of the basic numerical results, we conduct more ablation experiments in Fig. 11, which extends the semantic system to more transmission times, where $N$ represents the number of transmission times, and the maximum transmission time in this experiment is $N = 4$. 
The results prove that with the increase of transmission times $N$, the overall quality of transmission for both IK-HARQ schemes can be effectively improved.
Although the performance of IK-HARQ with a single decoder is a bit lower than the counterpart with multi-decoders at the same $N$, which is acceptable considering that it only uses single decoder with far less complexity.
Besides, with the increase of $N$, the performance of IK-HARQ with a single decoder (maximum $N = 4$) decreases slightly compared with IK-HARQ with a single decoder (maximum $N = 2$) when $N = 1, 2$, explained by the fact that it needs to deal with two additional situations ($N = 3, 4$).

Fig. 12 presents the performance about our proposed communication system  with adaptive semantic bit rate control.
It can be observed that the number of bits used for each sentence can significantly influence the BLEU score under different SNR.
Obviously, the more bits are used, the higher accuracy of transmission can be expected. Here, we adopt  three different bit settings (i.e., 30, 45 and 60). The performance of BLEU are consistent with our intuition.
More concretely,  the highest bit length (i.e., $B=60$) achieves the best performance, 
which is much better than the lowest bit length (i.e., $B = 30$).
Meanwhile, the performance gap between different bit lengths decreases with the increase of SNR.
There is almost no  performance difference when SNR is 6dB for $B = 45$ and $B = 60$.
Therefore we can take advantage of  flexible bit length on the basis of ensured accuracy, instead of fixed bit length. 
These experiments point out that with the help of a policy network, we can gain the capability to choose  appropriate
bit length according to the semantic characteristics of sentence and the channel SNR. 
The result of length selection can be observed from the pie chart in Fig. 13. To ensure the accuracy of transmission, we find the policy network tends to use higher bit length (i.e., $B$ = 45, 60) when  
channel condition is harsh (i.e., SNR = 0 dB).
With the improvement of SNR,  the length selection tends to choose 
lower length (i.e., $B$ = 30) for transmission resource saving on the premise of certain accuracy. 
The performance of the semantic communication system with adaptive semantic bit rate control can also be seen in Fig. 12, where the curve (\emph{Policy}) represents the corresponding BLEU performance.
As shown in Fig. 12, the two curves (i.e., curve ($B = 60$) and curve (\emph{Policy})) almost coincide with each other, which means that we can actually reduce the transmitted bit consumption without a significant performance drop. With the help of the policy network, the accuracy of transmission can reach that of using a higher fixed bit length (in this experiment is $B$ = 60); however compared with the fixed bit length,  the number of bits actually used can be greatly saved.

\begin{figure}[t]
	\includegraphics[scale=0.42]{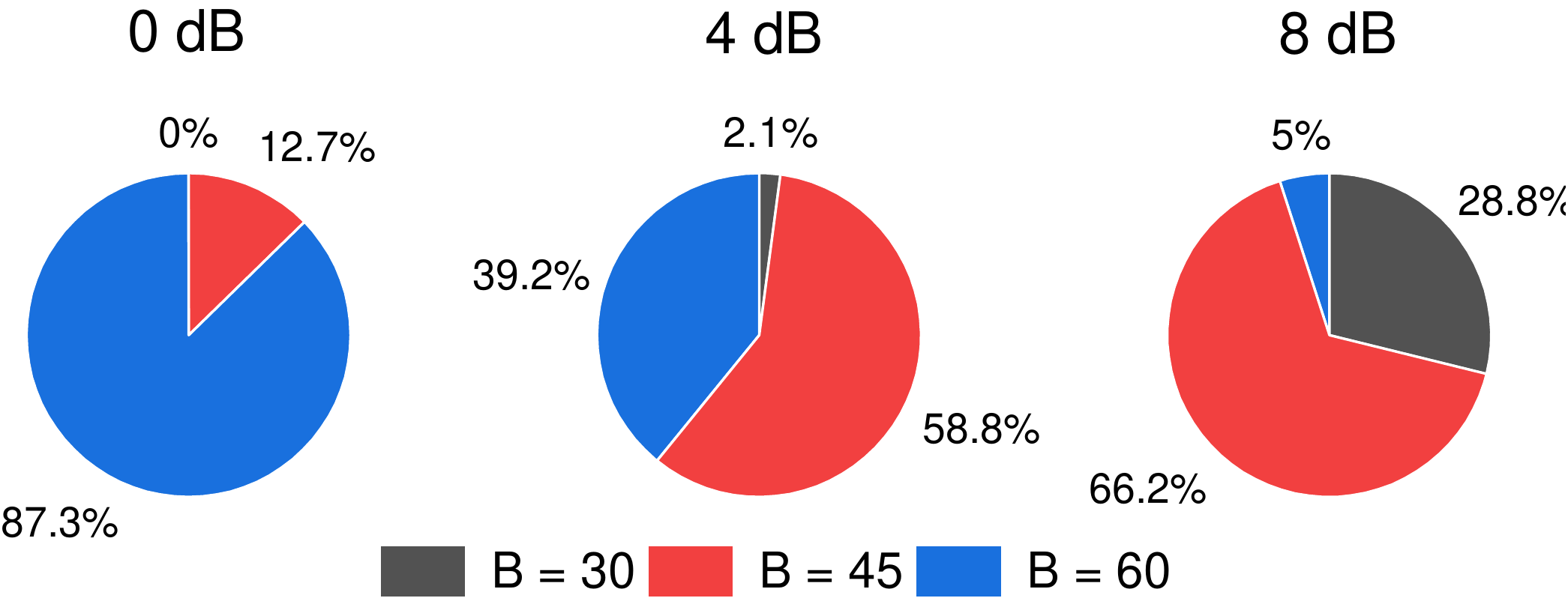}
	\caption{The distribution of the chosen bit length by the policy network.}
\end{figure}

The general performance of bit length adaptation with the policy network is shown in Fig. 14. 
In order to ensure a successful transmission, the policy network seeks to use more bits for each word when SNR is terrible. With the increase of SNR, the average number of bits used for each word generally shows a downward trend for both two curves.
Furthermore, we can notice that the curve (Denoising module) shows a better performance than the curve \emph{(No denoising module)}, since the entire system can acquire a desirable result with lower bit length at the same SNR after adding denoising modules.  
This also proves that the policy network is able to flexibly  adjust itself according to the actual performance of the semantic system.
Therefore, by comprehensively considering the semantic features and SNR, the policy network can help achieve minimum communication cost to accurately convey semantic information.

Fig. 15 shows the performance about semantic communication system with SNR-adaptive and self-adaptive denoising modules.
As we can observe from the figure, compared with the basic SC based scheme, the SC with SNR-adaptive and self-adaptive denoising modules achieve superior BLEU performance, respectively. 
By considering the transmission channel SNR, SNR-adaptive denoising module can further improve the performance of the semantic system.
When SNR = -4 dB, the BLEU performance can even be improved to almost 0.6. 
As SNR increases, the gap between basic SC and SC with SNR-adaptive module will gradually be narrowed. 
Different from the SNR-adaptive module, which needs to know the channel SNR,
there is no precondition for the self-adaptive denoising module. 
By exploiting  the semantic information between the contexts through self-attention layer,
the self-adaptive denoising module facilitates the intelligent scaling of semantic vectors and eventually leads to an adaptive self-denoising.
Although the gain is not as evident as SNR-adaptive denoising module, self-adaptive denoising requires no extra information, which is more practical in reality.
Meanwhile, when SNR = -4 dB, the BLEU performance can be improved to 0.35, which is also a satisfactory gain over the basic SC.
Similar to  SNR-adaptive denoising module, the gap between basic SC and SC with self-adaptive module will also gradually become smaller along with the improvement of SNR.
Finally, by combining these two denoising modules, the BLEU performance can be further improved, which shows the compatibility of these two denoising approaches.


\begin{figure*}[htbp]
\centering
\begin{minipage}{0.5\textwidth}
\centering
\includegraphics[scale=0.33]{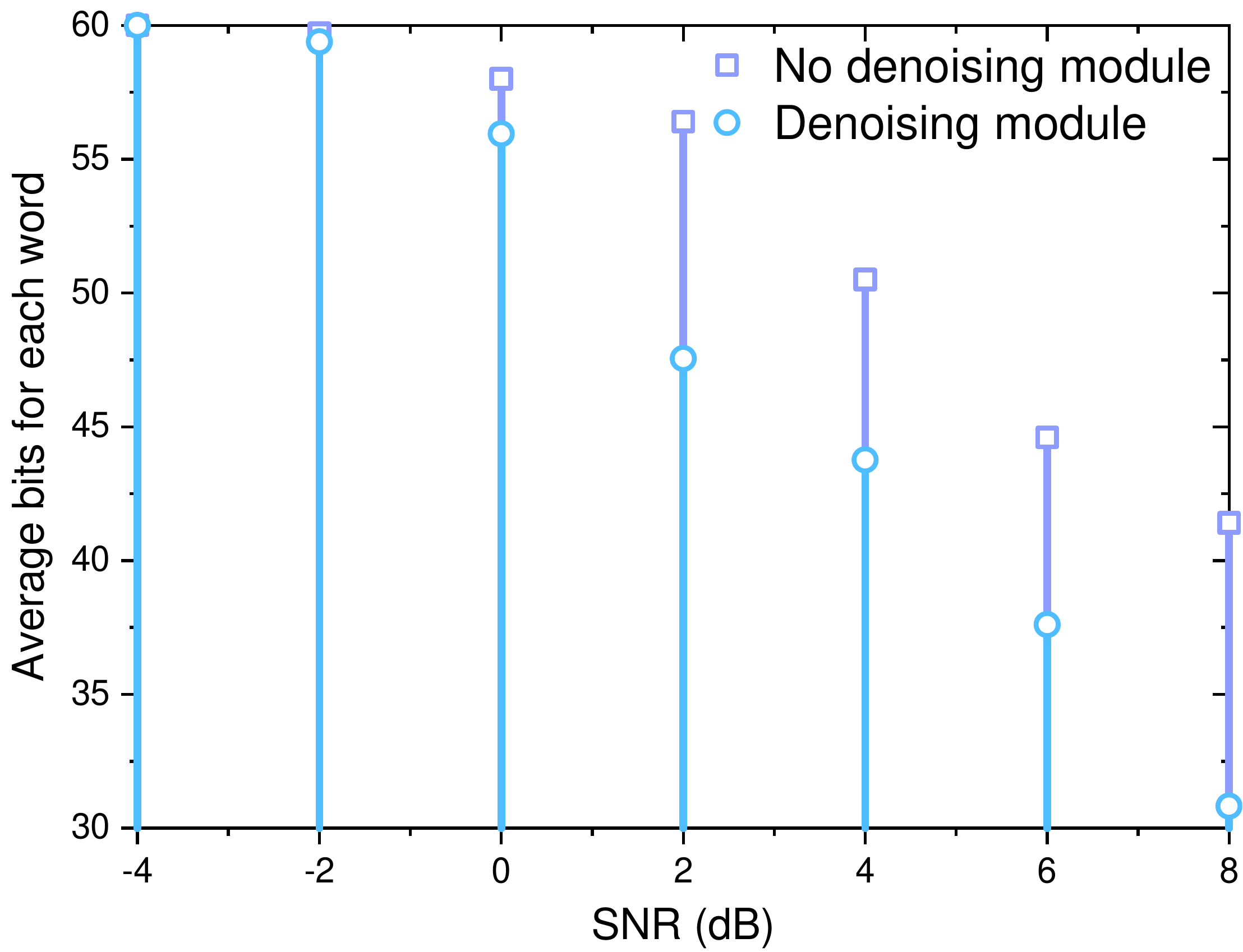}
\caption{Average bits for each word after using the policy network.}
\end{minipage}
\centering
\begin{minipage}{0.49\textwidth}
\centering
\includegraphics[scale=0.33]{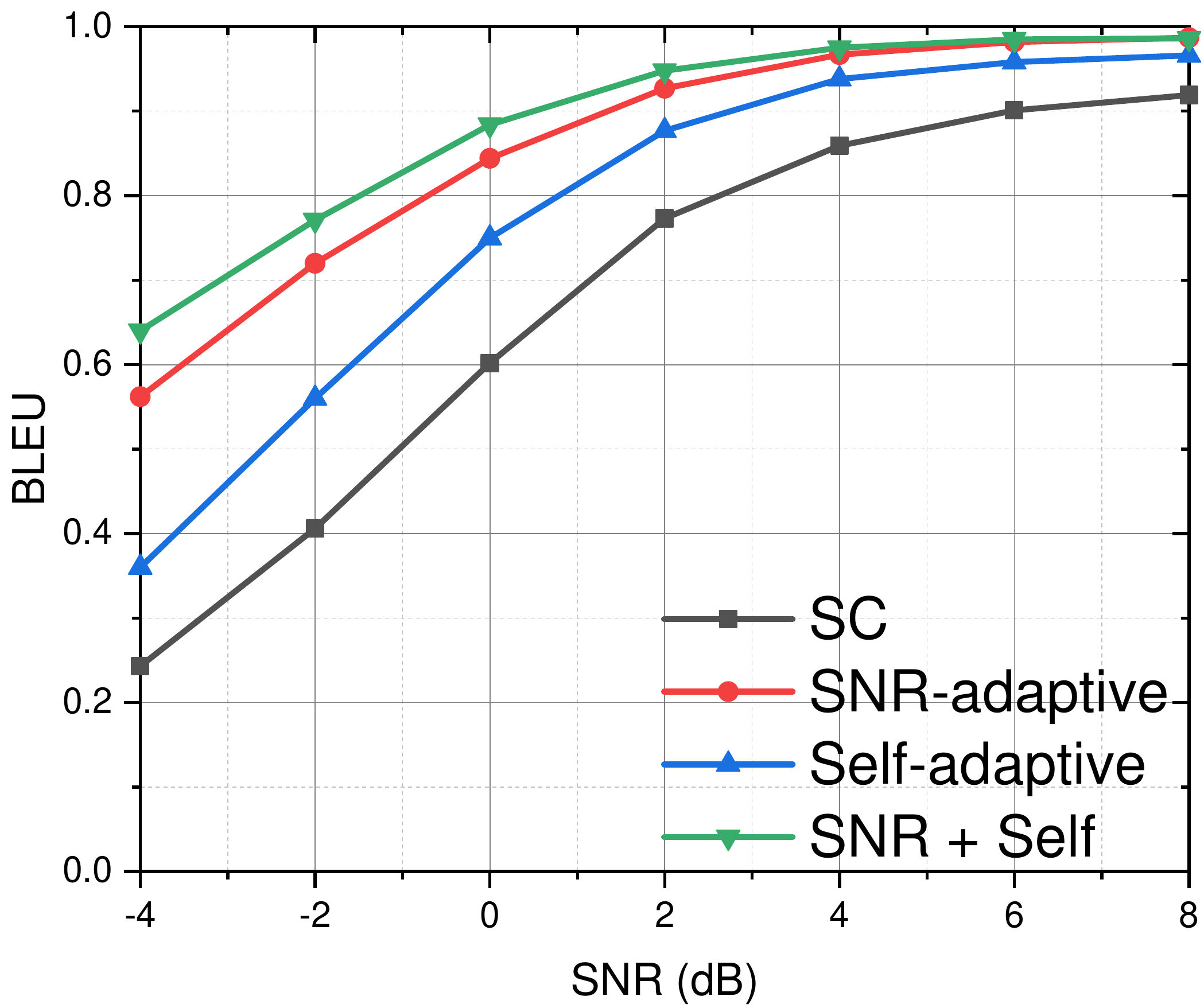}
\caption{BLEU score versus SNR for different denoising modules.}
\end{minipage}
\end{figure*}

\begin{figure}[t]
	\centering
	\includegraphics[scale=0.33]{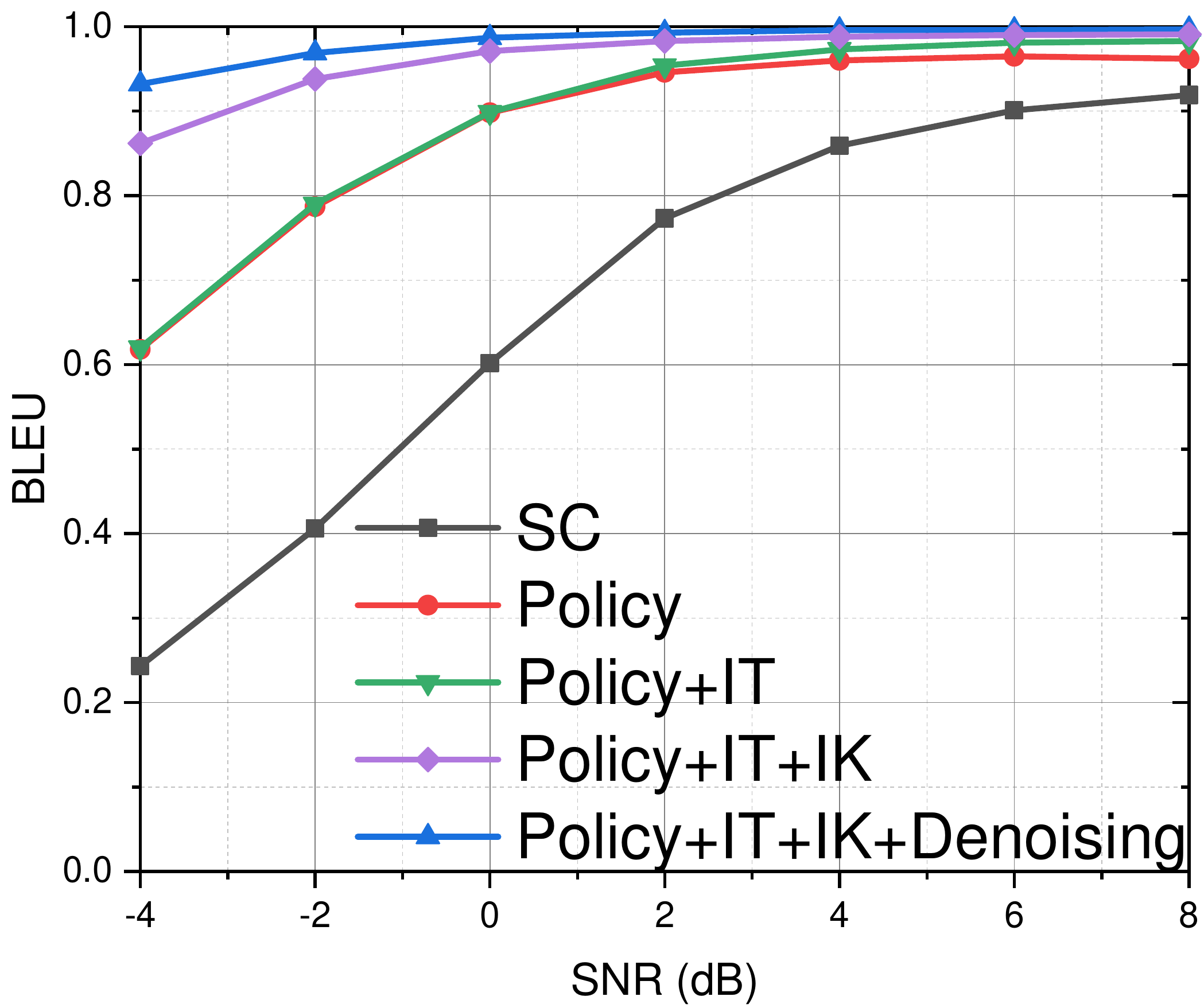}
	\caption{BLEU score versus SNR of the final integrated JSCC scheme.}
\end{figure}

The BLEU performance of our final integrated JSCC scheme is provided in Fig. 16. 
Here, we still adopt the same bit lengths (i.e., 30, 45 and 60).
The baseline  (i.e., curve \emph{(Basic SC)}) is the same as above under bit length \emph{B=30} without other modules.
The curve \emph{(Policy)} represents the variant using adaptive semantic rate control. 
Likewise, curve \emph{(Policy+IT)} means the encoder will retransmit incremental bits
if a transmission error is still encountered after the policy network selects the corresponding bit length (e.g. when the policy network chooses \emph{B=45} in this transmission, the encoder can transmit the remaining bits \emph{B=15} to enhance transmission performance).
While the curve \emph{(Policy+IT+IK)} shows BLEU performance when the entire system combines adaptive bit rate control, IT with IK-HARQ.
With the addition of denoising modules, the curve \emph{(Policy+IT+IK+Denoising)} displays the performance of the final integrated JSCC scheme.
Because the policy network tends to choose a higher bit length when the channel condition is relatively poor as shown in Fig. 14,  the curve \emph{(Policy)} and the curve \emph{(Policy+IT)} are almost coincident under low SNR, since there are no extra remaining bits. 
With the increase of SNR, the policy network tends to choose the lower bit length (i.e., $B$=30 or 45) which gives an opportunity for IT to further improve the BLEU performance.
Due to the characteristics of DNN, unlike traditional coding, we can not increase bit length in an unlimited manner once the structure of  semantic network has been determined.
However, due to the randomness of the wireless channel, transmission can not be guaranteed fully correct when even all the coding bits (i.e, $B$=60 in this experiment) are transmitted at one time.
Therefore, when there are no extra bits left, IK-HARQ could be leveraged to combat certain transmission errors. 
The curve (Policy+IT+IK) shows the intervention of extra incremental knowledge can significantly improve transmission accuracy when the bit length reaches the maximum threshold.
Besides, the curve \emph{(Policy+IT+IK+Denoising)} proves the effectiveness of our integrated scheme about denoising modules.

\begin{table}[t]
  \centering
  \caption{Complexity for each module}
\begin{threeparttable}
\begin{tabular}{c|c|c}
\hline
{Transmission system} & {Complexity} & {Maximum Path Length}\\ \hline
SC &  $O(L^{2} \cdot D)$ &  $O(L)$\\ \hline
IK-HARQ & $O(L^{2} \cdot D)$ &  $O(N \cdot L)$\\ \hline
Multi-Rate Coder & $O(L^{2} \cdot D)$ &  $O(L)$\\ \hline
Policy Network & $O(L \cdot D)$ &  $O(1)$\\ \hline
SNR-adaptive Denoiser & $O(L \cdot D)$ &  $O(1)$\\ \hline
Self-adaptive Denoiser & $O(L^{2} \cdot D)$ &  $O(1)$\\ \hline
\end{tabular}
\end{threeparttable}
\end{table}

\subsection{Implementation Cost and Computational Complexity}
One major concern of the proposed  approaches comes form the computational and implementation cost. Here, we analyse the complexity and maximum path length for each module as shown in Table III, 
where $L$ is the sentence length, $D$ is the embedding dimension , and $N$ represents the transmission times. Compared with the basic SC, IK-HARQ shows the same complexity $O(L^{2} \cdot D)$. Although there is one more step of information aggregation for IK-HARQ, the complexity of aggregation is  $O(L \cdot D)$, which is negligible compared with the complexity $O(L^{2} \cdot D)$, so the overall complexity is the same as SC. The maximum path length for IK-HARQ is affected by transmission times $N$ since the whole information needs to be decoded $N$ times.
The complexity and maximum path length for policy network and SNR-adaptive denoiser are $O(L \cdot D)$ and $O(1)$ respectively, considering the fact that they are composed of MLP.
It is worth noting  that the complexity of self-adaptive denoiser is $O(L^{2} \cdot D)$ because it is composed of self-attention layer. But different from SC whose maximum path length is $O(L)$, the maximum path length for self-adaptive denoiser is $O(1)$, considering the parallel operation for the self-attention mechanism.
So the use of a self-adaptive denoising module is equivalent to calculating a sentence with more words compared than the original, which is comparable with existing schemes and acceptable.

\section{Conclusion}
In this paper, we have explored the problem of adaptive bit rate control in  semantic communication, and proposed an end-to-end JSCC solution with IK-HARQ. 
In particular, for the encoder, we have put forward an adaptive semantic bit rate control encoder to jointly consider the impact of channel quality and semantic complexity of messages; while for the decoder, a semantic IK-HARQ is leveraged to enable the integrated decoding  of the latest transmission and previous failure experiences. Besides, a self-adaptive denoising module has been incorporated without requiring the SNR information as a prerequisite.

\bibliographystyle{IEEEtran}
\bibliography{IEEEabrv,reference. bib}
\nocite{*}

\end{document}